\documentclass[fleqn,usenatbib]{mnras}

\usepackage{newtxtext,newtxmath}

\usepackage[T1]{fontenc}
\usepackage{ae,aecompl}

\usepackage[utf8]{inputenc}
\usepackage{hyperref}

\usepackage{graphicx}	
\usepackage{amsmath}	
\usepackage{amssymb}	

\newcommand{\eagle}[0]{{\sc eagle}}
\newcommand{\illustris}[0]{{\sc Illustris}}
\newcommand{\illustrisTNG}[0]{{\sc IllustrisTNG}}
\newcommand{\GAMA}[0]{{\sc gama}}
\newcommand{\gadgettwo}[0]{{\sc Gadget-2}}
\newcommand{\gadgetthree}[0]{{\sc Gadget-3}}
\newcommand{\skirt}[0]{{\sc skirt}}
\newcommand{\sunrise}[0]{{\sc sunrise}}
\newcommand{\hyperion}[0]{{\sc hyperion}}
\newcommand{\GALAXEV}[0]{{\sc galaxev}}
\newcommand{\MAPPINGS}[0]{{\sc mappings-iii}}
\newcommand{\statmorph}[0]{{\sc statmorph}}
\newcommand{\MsunInline}[0]{{M$_\odot$}}
\newcommand{\swarp}[0]{{SWarp}}
\newcommand{\photutils}[0]{{\sc photutils}}

\title[]{Non-parametric Morphologies of Galaxies in the \eagle{} Simulation}

\author[Bignone et al.]{
Lucas A. Bignone,$^{1}$\thanks{E-mail: l.bignone@uandresbello.edu}
Susana E. Pedrosa,$^{2}$
James W. Trayford,$^{3}$
Patricia B. Tissera$^{1}$
\newauthor
and Leonardo J. Pellizza$^{4}$
\\
$^{1}$Departamento de Ciencias Físicas, Universidad Andrés Bello, Santiago, Chile \\
$^{2}$Instituto de Astronomía y Física del Espacio, CONICET-UBA, Casilla de Correos 67, Suc. 28, 1428, Buenos Aires, Argentina \\
$^{3}$Leiden Observatory, Leiden University, PO Box 9513, 2300 RA Leiden, the Netherlands \\
$^{4}$Instituto Argentino de Radioastronomía, (CICPBA – CONICET). Villa Elisa, Argentina
}

\date{Accepted -. Received -; in original form -}

\pubyear{2019}

\begin{document}
\label{firstpage}
\pagerange{\pageref{firstpage}--\pageref{lastpage}}
\maketitle

\begin{abstract}

 We study the optical morphology of galaxies in a large-scale hydrodynamic
 cosmological simulation, the \eagle{} simulation. Galaxy morphologies were
 characterized using non-parametric statistics (Gini, $M_{20}$, Concentration
 and Asymmetry) derived from mock images computed using a 3D radiative transfer
 technique and post-processed to approximate observational surveys. The
 resulting morphologies were contrasted to observational results from a sample
 of $\log_{10}(M_{*}/\textnormal{M}_\odot) > 10$ galaxies at $z \sim 0.05$ in
 the \GAMA{} survey. We find that the morphologies of \eagle{} galaxies
 reproduce observations, except for asymmetry values which are larger in
 the simulated galaxies. Additionally, we study the effect of spatial
 resolution in the computation of non-parametric morphologies, finding that Gini
 and Asymmetry values are systematically reduced with decreasing spatial
 resolution. Gini values for lower mass galaxies are especially affected.
 Comparing against other large scale simulations, the non-parametric statistics
 of \eagle{} galaxies largely agree with those found in \illustrisTNG{}.
 Additionally, \eagle{} galaxies mostly reproduce observed trends between
 morphology and star formation rate and galaxy size. Finally, We also find a
 significant correlation between optical and kinematic estimators of
 morphologies, although galaxy classification based on an optical or a kinematic
 criteria results in different galaxy subsets. The correlation between optical
 and kinematic morphologies is stronger in central galaxies than in satellites,
 indicating differences in morphological evolution.
 
 \end{abstract}

\begin{keywords}
methods: numerical – techniques: image processing – galaxies: formation – galaxies: statistics – galaxies: structure
\end{keywords}



\section{Introduction}

Galaxy morphology is not only important for classification, it also provides
crucial information on the evolution of galaxies. This is justified by the fact
that morphology is found to be strongly linked to the local environment
\citep{dressler_evolution_1984,gomez_galaxy_2003,blanton_physical_2009,kormendy_bulgeless_2010},
merger history \citep{lotz_galaxy_2008}, stellar mass
\citep{kauffmann_dependence_2003,ilbert_galaxy_2010} and star formation history
\citep{kauffmann_dependence_2003,baldry_quantifying_2004,bell_what_2012} of a
galaxy. See also \citet{conselice_evolution_2014} for a recent review on the
topic. In the last decade, large galaxy surveys have established the existence
of a bimodality in the nearby Universe where star-forming galaxies exhibit
disc-dominated (late-type) morphologies, while quiescent galaxies exhibit
bulge-dominated (early-type) morphologies. However, the detailed origin of the
observed distribution of morphologies is still debated, since the complete
physics of quenching and the assembly history of bulges is not yet fully
understood.

Numerical simulations offer unique insight into this problem because they can
link the morphology of a galaxy to the underlying physical processes that gave
rise to it in the first place. It is therefore desirable to be able to robustly
map the results of the simulations to morphological measurements that can be
contrasted to observations. A particularly powerful way to achieve this consists
in the generation and subsequent analysis of mock galaxy images from
hydrodynamic simulations. Several codes now exist that can produce such images:
\sunrise{} \citep{jonsson_sunrise_2006}, \hyperion{}
\citep{robitaille_hyperion_2011}, and \skirt{}
\citep{baes_efficient_2011,camps_skirt_2015}. These codes model the propagation
of photons trough the interstellar medium (ISM) and the effects of dust
absorption and scattering by solving the three-dimensional radiative transfer
calculations \citep[e.g.][]{steinacker_three-dimensional_2013} using Monte Carlo
techniques \citep[e.g.][]{whitney_monte_2011}

A key advantage of the use of mock images is that morphological analysis can be
performed in the same way in simulations and observations, using all of the
currently available techniques: non-parametric statistics
\citep[e.g.][]{lotz_galaxy_2008, snyder_galaxy_2015,
snyder_diverse_2015,bignone_non-parametric_2017,rodriguez-gomez_optical_2019};
bulge/disc decompositions based on S\'ersic indexes
\citep{scannapieco_observers_2010, bottrell_galaxies_2017}; human-based visual
classification \citep{dickinson_galaxy_2018} and machine learning algorithms
\citep{pearson_identifying_2019,huertas-company_hubble_2019}.

Non-parametric morphologies
\citep{lotz_new_2004,conselice_asymmetry_2000,freeman_new_2013,pawlik_shape_2016}
play a central role because they are generally more flexible than
classifications based on S\'ersic index, i.e. they can be used even in cases of
irregular morphologies \citep{lotz_evolution_2008}. Also, they are generally
easier to obtain, quantify and interpret than human- or machine- based visual
classification, although they do not provide as detailed morphological
taxonomies. All things considered, they provide a robust way to compare the
visual morphologies of observations and simulations.

In this work we study the optical morphologies of galaxies in the \eagle{}
simulations \citep{schaye_eagle_2015,crain_eagle_2015,mcalpine_eagle_2016} at $z=0.1$. To do so,
we use non-parametric statistics to quantify the light distribution of mock
galaxy images obtained using the radiative transfer code \skirt{}
\citep{camps_far-infrared_2016,trayford_optical_2017}, which include modelling
of dust absorption and scattering and that have been post-processed to mimic
SDSS and LSST images. We apply the same characterization techniques to SDSS
observations of galaxies in the \GAMA{} survey and compare our results. We also
contrast the morphologies of \eagle{} galaxies with that of other large-scale
simulations:
\illustris{} \citep{nelson_illustris_2015,vogelsberger_properties_2014} and \illustrisTNG{} 
\citep{nelson_illustristng_2019,pillepich_first_2018,springel_first_2018,nelson_first_2018,naiman_first_2018,marinacci_first_2018}. This work also complements other related studies based on the \eagle{}
simulations that characterize morphologies based on kinematic properties
\citep{correa_relation_2017,correa_origin_2019, lagos_connection_2018,
clauwens_three_2018, rosito_assembly_2018} or the combination of kinematics and
the spatial distribution of stellar mass
\citep{trayford_star_2019,thob_relationship_2019}.

This paper is organized as follows. In Section \ref{sec:simulation_and_data} we
describe the simulations used in this work and the simulated and observational
galaxy samples we characterize morphologically. In Section
\ref{sec:image_analysis} we describe the procedures used to obtain the simulated
and observational galaxy images and to derive the non-parametric statistics. We
present our main results in Section \ref{sec:results}. Finally, we summarize and
discuss these results in Section \ref{sec:Summary_and_discussion}.

\section{Simulation and data} \label{sec:simulation_and_data}

\subsection{The \eagle{} Simulations} \label{sec:the_eagle_simulations}

The \eagle{} simulations \citep{crain_eagle_2015, schaye_eagle_2015} are a suite
of cosmological hydrodynamic simulations run with a modified version of the
\gadgetthree{} N-Body Tree-PM smoothed particle hydrodynamics (SPH) code, which
is an updated version of \gadgettwo{} \citep{springel_cosmological_2005}. The
simulations follow the evolution of gas and dark matter in periodic cubic
volumes, with a range of resolutions and different parameter sets for the
subgrid models. The physics described by these subgrid models include the
heating and cooling of gas \citep{wiersma_effect_2009}, star formation
\citep{schaye_relation_2008}, stellar mass loss \citep{wiersma_chemical_2009},
energy feedback from star formation \citep{dalla_vecchia_simulating_2012} and
active galactic nuclei (AGN) feedback \citep{rosas-guevara_impact_2015}. The
model parameters regulating the energy feedback from star formation and AGNs
were calibrated so as to reproduced the observed galaxy stellar mass function
(GSMF) at $z \sim 0$. Additionally, a dependence of the stellar feedback energy
on the gas density was introduced so as to reproduce the galaxy mass-size
relation at $z \sim 0.1$. A comprehensive description of the calibration
procedure can be found in \citet{crain_eagle_2015}.

Star formation is treated stochastically in \eagle{} using a pressure-dependent
formulation of the empirical Kennicutt–Schmidt law following
\citep{schaye_relation_2008} but with a metallicity-dependent density threshold
\citep{schaye_star_2004}. Gas particles that are converted into stars inherit
the chemical abundance of their parent and are treated as single stellar
population, assuming a \citet{chabrier_galactic_2003} stellar initial mass
function (IMF). These stellar populations lose mass through stellar winds and
are subjected to mass loss via Type Ia and Type II supernovae events which
result in the chemical enrichment of their surrounding gas particles
\citep{wiersma_chemical_2009}. Abundances for eleven individual elements (H, He,
C, N, O, Ne, Mg, Si, S, Ca and Fe) are followed in the simulations.

In this work, we concentrate our analysis on the reference model (Ref) run in a
cosmological volume of 100 comoving Mpc on a side (Ref100N1504, hereafter
Ref-100). Additionally, in appendix \ref{sec:numerical_convergence} we analyse
two smaller volumes, 25 comoving Mpc a side (RefL025N0376 and RecalL025N0752),
to test the numerical convergence of non-parametric statistics. Key properties
of the \eagle{} simulations used in this work are listed in table
\ref{tab:eagle_sim_params}.

\begin{table}
	\centering
	\caption{Parameters of the \eagle{} and \illustris{} simulations used in this work. From left to
    right: simulation identifier, side length of cubic volume L in co-moving Mpc
    (cMpc), gas particle initial mass $m_\textnormal{g}$, Plummer equivalent gravitational
    softening $\epsilon_\textnormal{prop}$ at redshift z = 0 in proper kpc (pkpc)}
	\label{tab:eagle_sim_params}
    \begin{tabular}{l r r r r r r}
        \hline
        Name                        & $L$   & $m_\textnormal{g}$     & $\epsilon_\textnormal{prop}$ \\
                                    & cMpc  & $10^{5}$ \MsunInline{} & pkpc                         \\
        \hline                                                                                                                     
        Ref100N1504 (Ref-100)       & 100   & $18.1$                 & 0.70  \\
        RefL025N0376 (Ref-25)       & 25    & $18.1$                 & 0.70  \\ 
        RecalL025N0752 (Recal-25)   & 25    & $2.26$                 & 0.35  \\
        \illustris-1 (\illustris{}) & 106.5 & $12.6$                 & 0.71  \\
        TNG100-1 (\illustrisTNG{})                 & 110.7 & $13.9$                 & 0.74  \\
        \hline
        
    \end{tabular}
\end{table}

The cosmological parameters assumed by the \eagle{} simulations are those
inferred by the \citet{planck_collaboration_planck_2014}, the key parameters
being $\Omega_m = 0.307$, $\Omega_\Lambda = 0.693$, $\Omega_b = 0.04825$, $h =
0.6777$ and $\sigma_8 = 0.8288$. We adopt the same cosmological parameters for
this work.

\subsection{Simulated galaxy samples}
\label{sec:simulated_galaxy_samples}

\subsubsection{Ref-100}
\label{sec:Ref-100}

Dark matter haloes in \eagle{} are identified using the friends-of-friends (FoF)
algorithm \citep{davis_evolution_1985} with a linking length of $b=0.2$ times
the inter particle separation. Particles representing gas, stars and BHs are
associated with the FoF group of their nearest dark matter particle. Self-bound
substructures (subhalos) comprising dark matter, stars and gas are then
identified using the subfind algorithm \citep{springel_populating_2001}. Each
simulated galaxy is associated with an individual subhalo.

\begin{figure}
    \includegraphics[width=\columnwidth]{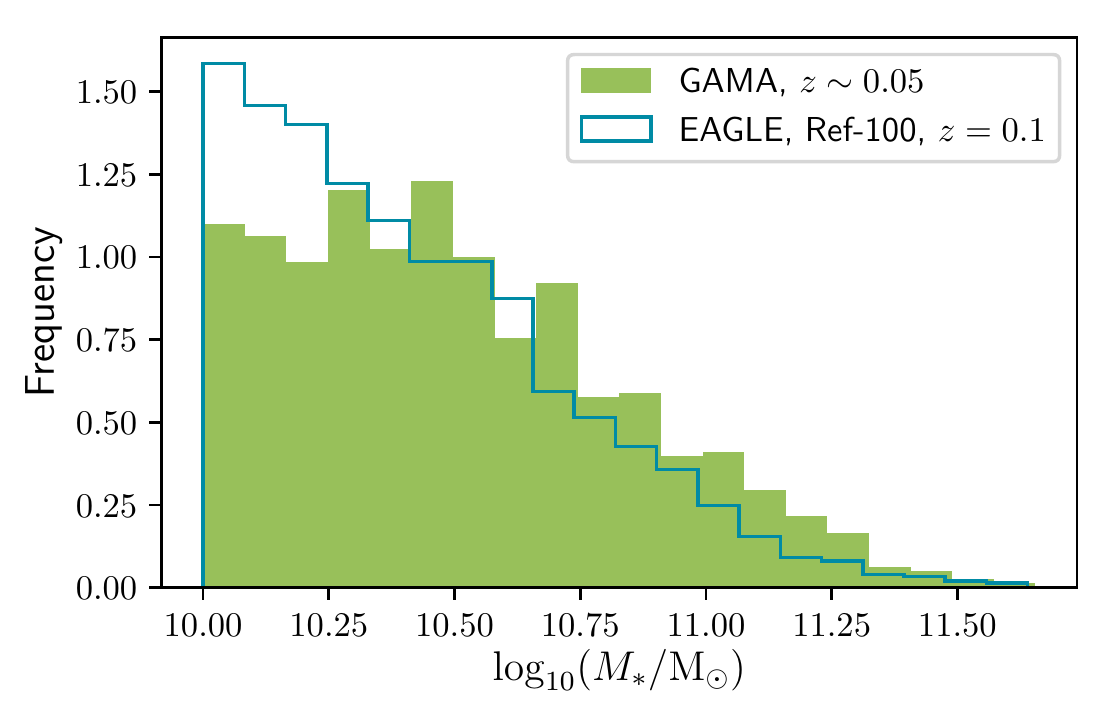}
    
    \caption{Stellar mass distribution of the \GAMA{} and Ref-100 samples. The
        \GAMA{} sample presents a slightly higher median stellar mass of
        $10^{10.45}$ \MsunInline{} compared to the median stellar mass of
        $10^{10.36}$ \MsunInline{} in the Ref-100 sample due to the paucity of
        galaxies below $\sim10^{10.5}$ \MsunInline{}. See text for a more
        complete discussion}
    
    \label{fig:samples_mass_distribution}
\end{figure}

We focus our study in galaxies in a single snapshot (snapshot 27) at $z = 0.1$
with $M_* > 10^{10}$ \MsunInline{}, the same galaxy sample for which images were
produced by \citet{trayford_optical_2017} using \skirt{}. The resulting mock
catalogue includes 3624 galaxies with most galaxies being resolved by more than
$10000$ stellar particles. The lowest number of stellar particle for a galaxy in
this sample is 6710. The sample contains 2255 (62\%) central and 1369 (38\%)
satellite galaxies. A summary of the properties of this simulated galaxy sample
can be found in table \ref{tab:galaxy_samples}.

Unless otherwise stated, all integrated galaxy properties (i.e. stellar mass,
star formation rate, half-light radius) in this sample are computed using
spherical apertures of 30 pkpc positioned on the centre of potential of the
corresponding galaxy.

\subsubsection{\illustris{} and \illustrisTNG{}}
\label{sec:illustris}

It is particularly interesting to compare our results with those obtained using
other cosmological simulations. This comparison can serve to illustrate
similarities and differences in the morphologies of simulated galaxies produced
by the different modelling of the physics of galaxy formation. Currently,
\illustris{} and \illustrisTNG{} are the two simulation suites that provide the
kind of non-parametric morphological studies that are directly comparable to
this work. 

Both \illustris{} and \illustrisTNG{} are a series of hydrodynamic cosmological
simulations run with the moving-mesh code AREPO, with \illustrisTNG{} featuring an updated
version of the Illustris galaxy formation model \citep{vogelsberger_model_2013,
torrey_model_2014}. The main ways in which \illustrisTNG{} differs from the original
Illustris are the inclusion of ideal magneto-hydrodynamics, a new AGN feedback
model that operates at low accretion rates \citep{weinberger_simulating_2017}
and modifications to the galactic winds, stellar evolution and chemical
enrichments according to \citet{pillepich_simulating_2018}.

In this study we use data from the highest resolution version of `TNG100',
hereafter \illustrisTNG{} and from the original `Illustris 1' simulation, hereafter
\illustris{}. Both simulations are very similar in terms of simulated volume and
resolution and differ mainly in the galaxy formation model. Basic Properties for
both simulations are detailed in Table \ref{tab:eagle_sim_params}.

We use non-parametric morphologies of galaxies extracted from
\citet{snyder_galaxy_2015} and from \citet{rodriguez-gomez_optical_2019}
corresponding to results from \illustris{} and \illustrisTNG{} respectively. Additionally,
for \illustris{}, we use asymmetries from \citet{bignone_non-parametric_2017}.
In all cases we impose a stellar mass threshold of $M_* > 10^{10}$
\MsunInline{}, matching the Ref-100 sample. This results in a sample of $7024$
($5926$) galaxies for \illustris{} (\illustrisTNG{}).

\subsection{The observational galaxy sample}
\label{sec:the_observational_galaxy_sample}

We consider galaxies in the Galaxy And Mass Assembly (\GAMA{}) survey
\citep{driver_gama_2009, robotham_galaxy_2010, driver_galaxy_2011}, a
spectroscopic and multiwavelength survey of five sky fields carried out using
the AAOmega multi-object spectrograph on the Anglo-Australian Telescope. The
survey has obtained $300000$ galaxy redshifts to $r < 19.8$ mag over $\sim286$
deg$^2$, with the survey design aimed at providing uniform spatial completeness.
The \GAMA{} survey provides us with a uniform galaxy database and a
comprehensive set of measure properties at low redshift that makes it ideal for
comparison with our simulated samples. 

\begin{table}
	\centering
	\caption{Properties of the simulated and observational galaxy samples used
        in this work: From left to right: designation, median stellar mass
        $\log_{10}$(<M$_*$>), redshift, number of galaxies in the sample N. For
        simulations, the listed redshifts represent the redshift from which the
        sample was extracted}
	\label{tab:galaxy_samples}
    \begin{tabular}{lccc}
        \hline
        Sample & $\log_{10}$(<M$_*$/M$_\odot$>) & Redshift & N\\
        \hline
        \eagle{} Ref-100 & ${10.36}$ & $0.1$ & 3624 \\
        \eagle{} Ref-25 & ${10.35}$ & $0.1$ & 70 \\
        \eagle{} Recal-25 & ${10.26}$ & $0.1$ & 75 \\
        \illustris{} & ${10.39}$ & $0.0$ & 7024 \\
        \illustrisTNG{} & ${10.43}$ & $0.05$ & 5926 \\
        \GAMA{} & ${10.45}$ & $0.045 < z < 0.06$ & 944 \\
        
        \hline
    \end{tabular}    
\end{table}

Comparisons between \eagle{} and \GAMA{} galaxies have been carried out in
several occasions, including during the calibration procedure mentioned in
section \ref{sec:the_eagle_simulations}, where the observed GSMF and size–mass
relation at $z \sim 0.1$ was used to determine the feedback model parameters
\citep{schaye_eagle_2015,crain_eagle_2015}. This means that neither the GSMF nor
the galaxy sizes can be presented as predictions of the simulations at this
redshift. The morphologies of galaxies have not been taken into consideration on
the calibration procedure and can therefore be contrasted fairly with \GAMA{},
or other similar galaxy samples to determine if the simulation reproduces
morphologies and what factors contribute to the establishment of optical
morphology. Previously, \eagle{} galaxy colours from mock \skirt{} images were
also compared with the \GAMA{} colour-mass diagram by
\citet{trayford_optical_2017}. They found that optical SKIRT galaxy colours
matched observations well and that the modelling by \skirt{} of the scattering
and absorption effects of dust improved the agreement with observations,
compared to more simple dust-screen models \citep{trayford_colours_2015}.

In this work, we make use of a galaxy subsample derived from the three \GAMA{}
equatorial fields, dubbed G09, G12 and G15, and restricted in redshift and
stellar mass. The photometrically derived stellar mass estimates are taken from
version 19 of the \GAMA{} stellar mass catalogue (internal designation {\sc
StellarMasses}) which were computed according to \citet{taylor_galaxy_2011},
corrected for aperture and re-scaled to the \eagle{} cosmology. Spectroscopic
redshifts are provided by the {\sc StellarMasses} catalogue. We restrict our
sample to galaxies with stellar mass $M_* > 10^{10}$ \MsunInline{} to match the
simulated mass range (Section \ref{sec:simulated_galaxy_samples}) and with
redshifts in the interval $0.045 < z < 0.06$ (median redshift $\sim 0.05$),
resulting in a total of 944 galaxies. The narrow redshift band allows us to
compare morphologies in the observed and simulated samples without considering
the effects of evolution in the galaxy population. The median redshift at 0.05
was chosen to match previous works on simulated galaxy morphologies
\citep{snyder_galaxy_2015, bignone_non-parametric_2017, dickinson_galaxy_2018,
rodriguez-gomez_optical_2019}. Additionally, this redshift represents the limit
at which the resolution of SDSS imaging starts to have a significant impact on
the reliability of standard non-parametric morphologies (we analyse image
resolution effects in more detail in section \ref{sec:resolution_effects}). A
summary of the properties of this observational sample can be found in Table
\ref{tab:galaxy_samples}.

Figure \ref{fig:samples_mass_distribution} compares the stellar mass
distribution of galaxies in the simulated and observational samples. It
illustrates the similarities and differences between both galaxy populations. A
flattening around and below $10^{10.5}$ \MsunInline{} can be appreciated in the
\GAMA{} sample which corresponds to the same feature in the GSMF discussed in
\citet{baldry_galaxy_2012}. While Ref-100 does not show a similar flattening,
the general shape of the GSMF agrees with observations to $\lesssim 0.2$ dex for
the full mass range for which the simulation resolution is adequate, i.e. from
$2 \times 10^8$ \MsunInline{} to over $10^{11}$ \MsunInline{}
\citep{schaye_eagle_2015}. Given that uncertainties in the stellar evolution
models used to infer stellar masses are $\sim0.3$ dex
\citep[e.g.][]{conroy_propagation_2009}, we can consider that the distribution
of stellar masses in both samples are comparable for the purposes of this work.
The paucity of galaxies towards the lower end of the mass range in the \GAMA{}
sample results in a slightly higher median mass of $10^{10.45}$ \MsunInline{},
compared to the median stellar mass of $10^{10.36}$ \MsunInline{} in the Ref-100
sample.

We obtain morphological information for our observational sample by cross
referencing all objects with the catalogue presented by
\citet{dominguez_sanchez_improving_2018}. This catalogue provides morphologies
for $\sim 600000$ galaxies based in the T-Type classification
\citep{de_vaucouleurs_revised_1963} and in the Galaxy Zoo 2 (GZ2) classification
scheme. To achieve that task, they combined existing visual classification
catalogues with Convolutional Neural Networks (CNNs) achieving $>97\%$ accuracy
for GZ2 morphologies, as well as no offset and a scatter comparable to typical
expert visual classifications for T-type morphologies.

\section{Image Analysis}
\label{sec:image_analysis}

\subsection{Simulated galaxy images}
\label{sec:simulated_galaxy_images}

For galaxies in our simulated sample, we utilize the mock images presented in
\citet{trayford_optical_2017} and generated using the radiative transfer code
\skirt{}. Here, we summarize the most relevant aspects of the image generation
procedure, but interested readers are recommended to refer to the original paper
for details.

The \skirt{} Monte Carlo code works by computing the absorption and scattering
of monochromatic photon packets from their origin at luminous sources to their
destination at a user-defined detector. It is possible to define imaging
detectors with a set distance from the source, field of view (FOV) and number of
pixels. Datacubes are produced by adding the flux at the position of each pixel
separately for each of the wavelengths sampled by the photon packets. Broadband
images can then be constructed by convolving the datacubes with the desired
filters. 

In this paper, we only consider the mass distribution associated with individual
subhalos (either centrals or satellites), leaving out close companions and other
members of the same halo. This makes the determination of morphologies for
individual galaxies robust. The effect of contamination from close companions or
background and foreground galaxies are not included in the mock images. Finally,
only stellar and gas particles within 30 pkpc of the galaxy centre are
considered, a choice initially made in \citet{trayford_colours_2015} to
reasonably approximate a Petrosian aperture, but which leaves out some of the
light distribution at the outskirts of the most extended galaxies.

\subsubsection{Photon sources}

Star particles representing stellar populations are used as the sources of the
photon packets. The number of photons in each wavelength of the spectral grid is
determined by assuming a spectral energy distribution (SED). There are different
types of SEDs assigned depending on stellar age. Old stellar populations (age
$>10$ Myr) are assigned a \GALAXEV{} \citep{bruzual_stellar_2003} SED as
described in \citet{trayford_colours_2015} and assumed to have a
\citet{chabrier_galactic_2003} IMF in the [$0.1 - 100$] \MsunInline{} mass
range. Young stellar populations (age $<10$ Myr) are treated differently because
the inability of the simulation to resolve the sub-kpc birth clouds were these
stars are embedded. For these stars, the \MAPPINGS{} spectral models of
\citet{groves_modeling_2008} are used, which include dust absorption within the
photodissociation region (PDR). Additionally, a re-sampling of stellar and
star-forming gas particles is carried out to to mitigate the effects of coarse
sampling due to the limited mass resolution \citep[similar
to][]{trayford_colours_2015}. Under this procedure, recent star formation is
re-sampled in time over the past 100 Myr. Stellar populations re-sampled with
ages younger than $10$ Myr are treated with the \MAPPINGS{} spectral models,
while those with ages older than $>10$ Myr with the \GALAXEV{} models.

The point of emission of individual photons is determined by randomly sampling
truncated Gaussian distributions centred at the position of stellar sources and
characterized by a smoothing length. This serves to represent the fact that
particles in the simulation do not correspond to individual point sources, but
mass distributions instead. Here the distance to the 64th nearest neighbouring
star was used as the smoothing length \citep[similar
to][]{torrey_synthetic_2015}. In general terms, the choice of smoothing length
has an impact on the appearance of the images, resulting in excessive
granularity or oversmoothing and therefore, can have an impact on non-parametric
morphologies.

\begin{figure*}
    \includegraphics[width=\textwidth]{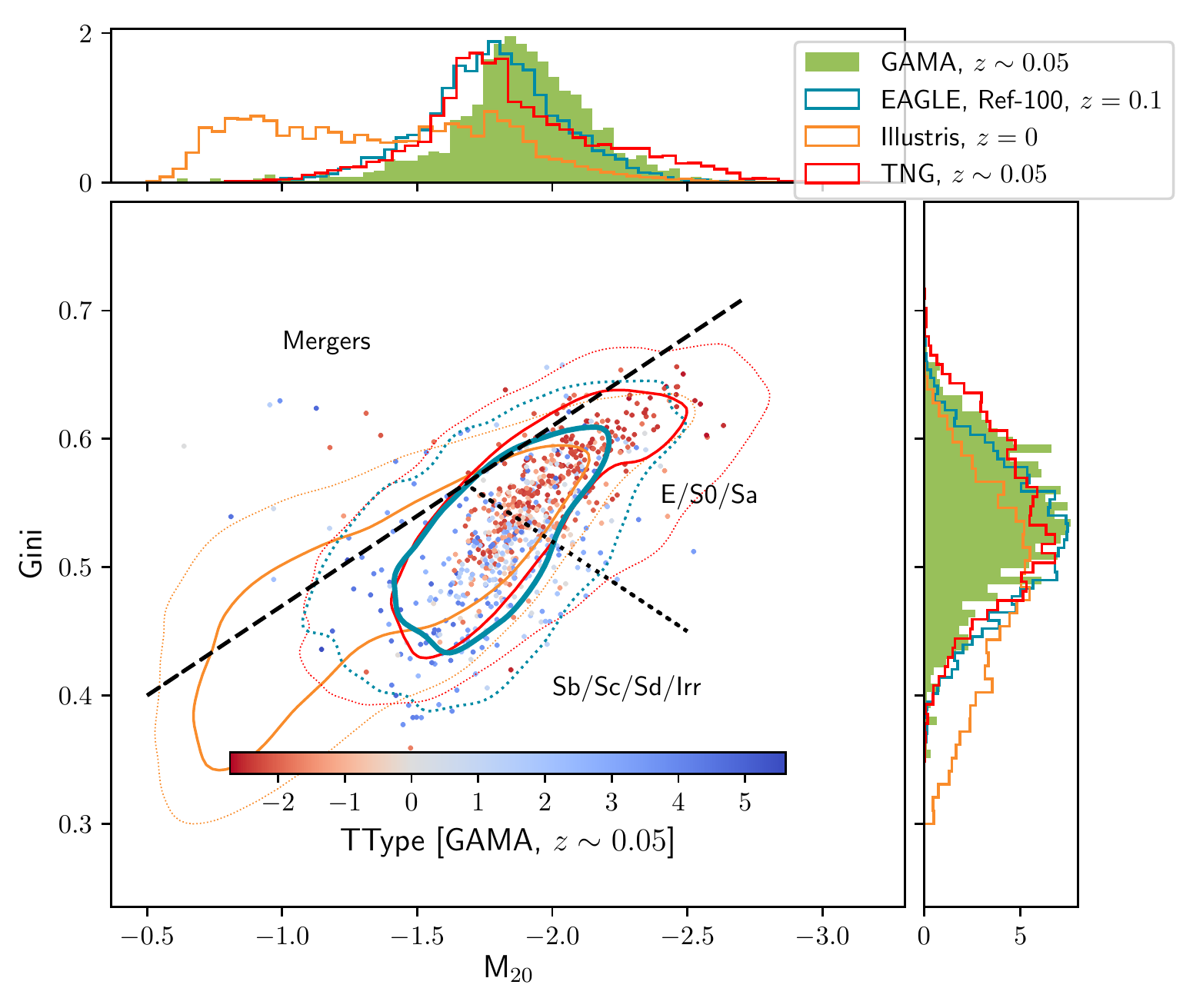}
    
    \caption{The central panel shows the $G$-M$_{20}$ diagram from galaxies in
        Ref-100 (blue), \illustris{} (orange), \illustrisTNG{} (red) and
        \GAMA{} (points). The coloured solid (dotted) lines enclose regions
        containing 68 (95) percent of galaxies in each respective sample. The
        \GAMA{} galaxies are coloured according to their T-Types. The black
        dashed and dotted lines separate the subspace into regions for mergers,
        late types and early types according to \citet{lotz_evolution_2008}. We
        find that Ref-100 and \illustrisTNG{} have very similar distributions
        in $G$-M$_{20}$ space and that they both match \GAMA{} observations.
        The top and right panels show respectively the $G$ and M$_{20}$
        normalized distributions for all samples.}
    
    \label{fig:gini_m20_scatter}
\end{figure*}

\subsubsection{Dust modelling}

Dust can have an important impact on the appearance of galaxies and is therefore
important to account for its effects. In this work, the distribution of dust in
the diffuse ISM is approximated by the distribution of gas within the
simulation. Since dust is observed to trace the cold metal-rich gas in observed
galaxies \citep[e.g.][]{bourne_herschel-atlas_2013}, a constant dust-to-metal
mass ratio is assumed \citep{camps_far-infrared_2016}
\begin{equation}
    f_\textnormal{dust} = \frac{\rho_\textnormal{dust}}{Z \rho_g} = 0.3,
\end{equation}
where $Z$ is the (SPH-smoothed) metallicity, and $\rho_\textnormal{dust}$ and
$\rho_g$ are the dust and gas density, respectively. Only non
star-forming and cold ($T < 8000 K$) gas contributes to the dust budget.

The dust density is mapped to an adaptively refined (AMR) grid with a minimum
cell size of $0.11$ kpc, close to the spatial resolution of \eagle{}. \skirt{}
then computes optical depths of each cell at a given reference wavelength and
the resulting obscuration. Dust composition is assumed to follow the model
described by \citet{zubko_interstellar_2004}; a multicomponent dust mix tuned to
reproduce the abundance, extinction and emission constraints of the Milky Way.

\subsubsection{Realistic images}
\label{sec:realistic_images}

The initial datacubes produced for our simulated galaxy sample span $256 \times
256$ spatial pixels and 333 wavelengths in the range $0.28-2.5$ $\mu$m, chosen
to sample the rest-frame ugrizYJHK photometric bands. Each datacube slice covers
a $60 \times 60$ kpc area. The camera location is set at 10 Mpc from the galaxy,
which results in a pixel scale of $\sim 235$ pc, sufficiently small to simulate
SDSS and LSST images for sources at $z > 0.02$. The images correspond to a
random orientation with respect to the galaxy (but fixed to the $xy$ plane of
the simulation box)\footnote{The effect of galaxy orientation on the
non-parametric statistics is explored in appendix
\ref{sec:dependence_on_rotation}}.

We concentrate our morphological analysis on rest-frame broadband g-band, SDSS
images obtained by convolving the datacubes in the wavelength dimension with the
corresponding filter transmission curve \citep{doi_photometric_2010}.

We then follow a procedure very similar to that of \citet{snyder_galaxy_2015} to
transform the noise-free, ideal images, into realistic images comparable to
SDSS observations at $z \sim 0.05$. The procedure can be summarized as follows:
\begin{itemize}
    \item We first convolve each image with a Gaussian point-spread function
    (PSF) with a full width at half-maximum (FWHM) of 1 kpc. At $z=0.05$ this
    approximates the effect of a 1 arcsec seeing, which roughly matches that of
    SDSS. Alternatively, the resulting mock images can also be comparable to
    more distant HST imaging at $z=0.5$. We explore other values for the FWHM to
    study the effect of seeing on non-parametric morphologies in Section
    \ref{sec:resolution_effects}.

    \item Next, we rebin the image to a constant pixel scale of $0.24$
    kpc pixel$^{-1}$, which again, roughly matches SDSS imaging. 

    \item Finally, we add Gaussian noise to the images such that the
    average signal-to-noise ratio of each galaxy pixel is 25. Therefore,
    we simulate only strongly detected galaxies with morphological
    measurements not affected by noise.
\end{itemize}

Also shown through this paper, for illustration purposes, are three-colour gri
images based on the ugriz SDSS bands and computed via the approach of
\citet{lupton_preparing_2004}. These images correspond to those publicly
available in the \eagle{} database \citep{mcalpine_eagle_2016} and have not been
degraded in the manner described above.

\subsection{Observational sample images}

For each galaxy in our \GAMA{} sample, we downloaded g-band SDSS images from the
online Data Release 12 (DR12)
archive\footnote{\url{https://www.sdss.org/dr12/imaging/images/}}. We made use
of the mosaic tool\footnote{\url{https://dr12.sdss.org/mosaics}} and the
\swarp{} tool \citep{bertin_terapix_2002} to obtain images centred at the
position of the object and restricted to an area $60 \times 60$ kpc at the
corresponding redshift, matching the limit imposed in our mock images.

The images are sky-subtracted and have a constant pixel scale of $0.396$ arcsec
pixel$^{-1}$, which is equivalent to $\sim0.396$ kpc pixel$^{-1}$ at the median
$z=0.05$ redshift of the sample.

\subsection{Structural measurements}

To compute non-parametric morphologies of both simulated and observational
samples, we use \statmorph{}, a Python package especially developed for this task
and used to compute optical morphologies of galaxies in the IllustrisTNG
simulation \citep{rodriguez-gomez_optical_2019}. We concentrate on the
computation of Gini ($G$), M$_{20}$, Concentration ($C$) and Asymmetry ($A$), although the code
also allows for the determination of additional morphological parameters.

Details regarding the specific computation of the non-parametric morphologies
can be found in \citet{rodriguez-gomez_optical_2019}, the implementation is
largely based on \citet{lotz_new_2004} for the case of $G$-M$_{20}$ and
\citet{conselice_relationship_2003} for $C$ and $A$. Here we
give a brief summary of how each statistic is measured

\subsubsection{Gini}

The Gini coefficient is a statistical tool that measures the distribution of a
quantity among a population, In the case of galaxy structure, it measures the
distribution of light among the pixels that encompass the galaxy image
\citep{lotz_new_2004}; higher values indicate a very unequal distribution (light
is mostly concentrated in a few pixels), whereas a lower value indicates a more
even distribution. The value of $G$ is defined by the Lorentz curve of the
galaxy’s light distribution according to 

\begin{equation}
    \textnormal{G} = \frac{1} {|\bar{f}| n(n-1)} \sum_i^n{(2i - n - 1) f_i},
\end{equation}

where $f_i$ are a set of $n$ pixel flux values, $i$ ranges from 0 to $n$ and
$\bar{f}$ is the average pixel flux value. At the extremes, a value of $G$ = 1
is obtained when all of the flux is concentrated in a single pixel, while $G$ =
0 results from a totally homogeneous flux distribution.

\subsubsection{$M_{20}$}

The second-order moment parameter, $M_{20}$ gives a value that indicates whether
light is concentrated within an image. However, unlike the $C$
statistic, which we define later, $M_{20}$ does not necessarily imply a central
concentration. Instead, light could be concentrated in any location in a galaxy.
Specifically, the value of $M_{20}$ is the moment of the fluxes of the brightest
20 per cent of light in a galaxy, which is then normalized by the total light
moment for all pixels ($M_\textnormal{tot}$) \citep{lotz_new_2004}.
$M_\textnormal{tot}$ is given by
\begin{equation}
    M_\textnormal{tot} = \sum_i^n {M_i} = \sum_i^n {f_i[(x_i - x_c)^2 + (y_i - y_c)^2]},
\end{equation}
where $f_i$ are the pixel flux values and ($x_c$, $y_c$) is the galaxy’s centre.

$M_{20}$ is then obtained by sorting the pixels by flux and summing $M_i$ over
the brightest pixels until the sum of the brightest pixels equals 20 per cent of
the galaxy’s total flux
\begin{equation}
    M_{20} = \log_{10}{\frac{\sum{M_i}}{M_\textnormal{tot}}}, \textnormal{while} \sum_i{f_i} < 0.2 f_n
\end{equation}

\subsubsection{Concentration}

The $C$ statistic quantifies how much light is in the centre of a
galaxy as opposed to its outer parts. It is usually defined
\citep{conselice_asymmetry_2000} as  
\begin{equation}
    \textnormal{$C$} = 5 \times \log_{10}{ \frac{r_{80}} {r_{20} } },
\end{equation}
where $r_{20}$ and $r_{80}$ are the radii of apertures containing 20 and 80 per
cent of the total flux, respectively. In the implementation of \statmorph{}, the
total flux is measured within a 1.5 petrosian radius and the centre of the
aperture corresponds to the point that minimizes the $A$ index.

\subsubsection{Asymmetry}
\label{sec:code_asymmetry}

Asymmetry is obtained by subtracting the galaxy image rotated by 180$^{\circ}$
from the original image \citep{conselice_asymmetry_2000}. It is given by
\begin{equation}
    \textnormal{$A$} = \frac{ \sum_{i,j} {|f_{ij} - f_{ij}^{180}|} } { \sum_{i,j} {|f_{ij|} } } - A_{0},
\end{equation}
where $f_{ij}$ and $f_{ij}^{180}$ are the pixel flux values of the original and
rotated image respectively, and $A_{0}$ is an estimation of the background
asymmetry. The sum is carried out over all pixels within 1.5 petrosian radius of
the galaxy’s centre, which is determined by minimizing $A$.

In the original implementation of \statmorph{}
\citep{rodriguez-gomez_optical_2019}, $A_{0}$ is computed as the average
asymmetry of the background. However, for galaxies that have a very symmetric
light distribution, or alternatively, where the S/N is low, the $A$ value can
become dominated by the sky background asymmetry average. This result in
artificially low and even negative asymmetry values. To compensate this, we
modified the code slightly so that the background asymmetry is instead computed
using a centroid pixel that minimizes its value. This is similar to the
procedure described by \citet{conselice_asymmetry_2000} and also implemented by
\citet{bignone_non-parametric_2017} for \illustris{} galaxies. It results in
mostly positive asymmetry values, shifted about 0.05 dex higher with respect to
the original \statmorph{} code.

\subsubsection{Segmentation maps}

In order to perform the morphological measurements, an initial segmentation map
that determines which pixels belong to the galaxy of interest is required. To
create the segmentation maps, we utilize the \photutils{} photometry
package\footnote{\url{https://photutils.readthedocs.io}}. For the mock sample we
find robust segmentation maps by setting the detection threshold at $1.2\sigma$
above the sky median, with the background level computed by \photutils{} using
simple sigma-clipped statistics. For the observational sample, there is the
significant problem of source contamination, therefore we apply an additional
deblending step using the {\sc deblend\_sources} routine, which uses a
combination of multi-thresholding and watershed segmentation to isolate sources.
In all cases, we only keep the source detected at the centre of the image, since
by construction, it must correspond to the object of interest. A final visual
inspection ensures that segmentation maps are reasonable, and that clumpy
star-forming galaxies in particular are not artificially fragmented. We find
that no manual corrections are necessary.

From this point on, both observational and simulated samples are processed by
\statmorph{} in the exact same manner to compute their respective non-parametric
morphologies.

\section{Results}
\label{sec:results}

\begin{figure*}
    \includegraphics[width=\textwidth]{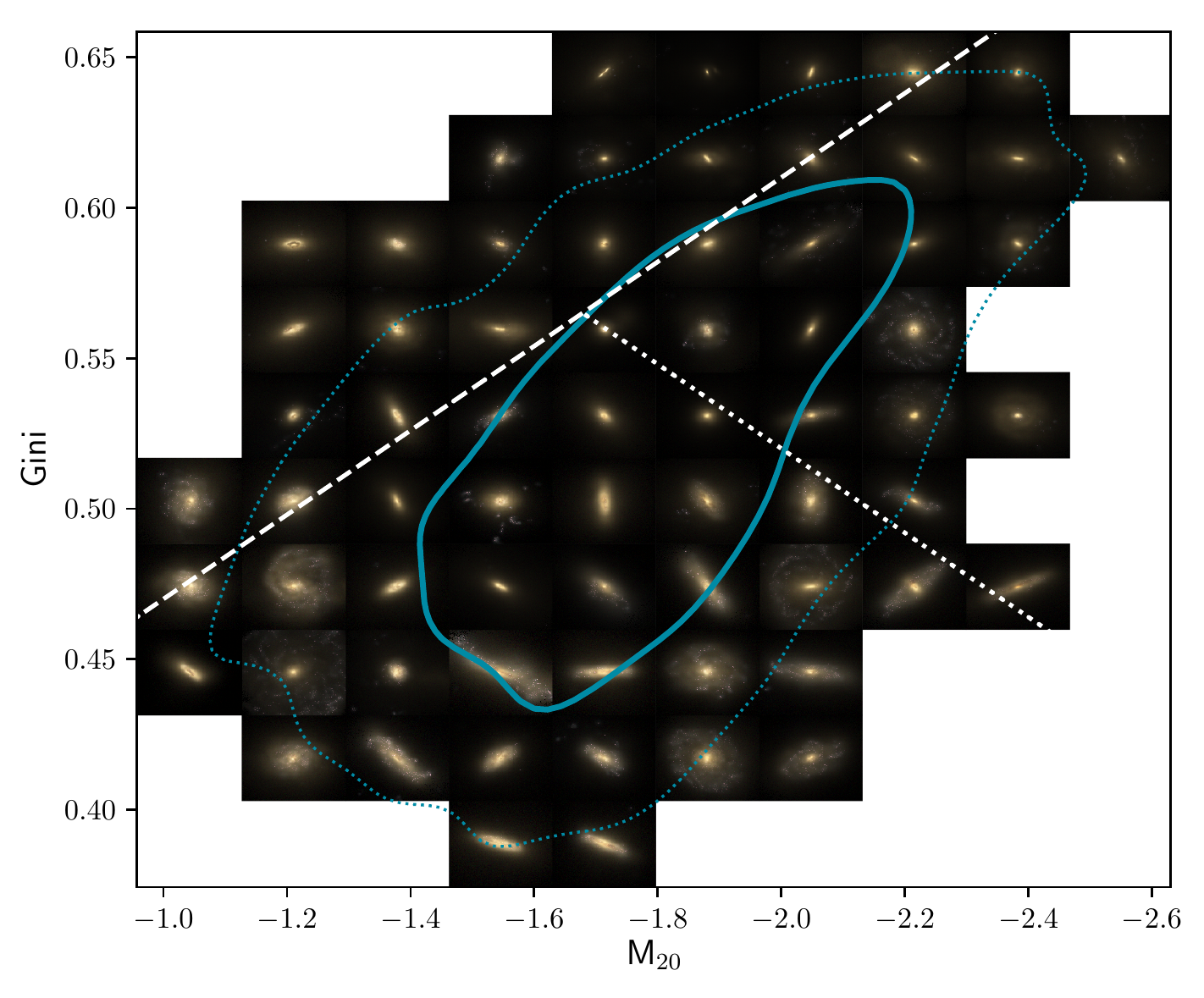}
    
    \caption{Mock gri-SDSS colour composite images of galaxies in the Ref-100 sample
    arranged according to their $G$ and M$_{20}$ values. Solid (dotted)
    contours represent the region containing 68 (95) percent of objects. The
    straight lines are as in Figure \ref{fig:gini_m20_scatter}. In accordance
    with observational trends, prevailing morphologies at the top right of the
    diagram are of the early type, while galaxies at the lower left are
    late-type. Signs of disturbed morphologies can be found above the dashed
    line for some galaxies.}
    
    \label{fig:G_M20_images}
\end{figure*}

\subsection{Gini-M$_{20}$}

Figure \ref{fig:gini_m20_scatter} shows the position in the $G$-M$_{20}$
morphological subspace occupied by the Ref-100 sample (blue contours), the
\illustris{} sample (orange contours), the \illustrisTNG{} sample (red) and the
GAMA sample (coloured points). The subspace is divided into three sectors where,
according to \citep{lotz_evolution_2008}, galaxies in the Extended Groth Strip
at $0.2 < z < 1.2$ present the following distinct morphologies:
\begin{description}
    \item [Mergers:] $\textnormal{G} > -0.14 \textnormal{M}_{20} + 0.33$,
    \item [E/S0/Sa:] $\textnormal{G} \leq -0.14 \textnormal{M}_{20} + 0.33$ and $\textnormal{G} > 0.14 \textnormal{M}_{20} + 0.80$,
    \item [Sb--Irr:] $\textnormal{G} \leq -0.14 \textnormal{M}_{20} + 0.33$ and $\textnormal{G} \leq 0.14 \textnormal{M}_{20} + 0.80$.
\end{description}
Galaxies in the \GAMA{} sample are colour coded according to the T-Type assigned
to them by the machine learning algorithm of
\citet{dominguez_sanchez_improving_2018}. It is clear that galaxies with
negative T-Type (corresponding to early-type galaxies) and those with with
positive T-Types (corresponding to late-type galaxies) prefer different
locations in the $G$-M$_{20}$ plane and that their positions generally agree
well with those determined by \citet{lotz_evolution_2008} for their respective
morphological type, with some intermixing.

It can also be appreciated in Fig. \ref{fig:gini_m20_scatter} that the location
occupied by galaxies in the Ref-100 (blue contours) coincides to a large extent
with that of the \GAMA{} sample. This constitutes strong evidence that the
morphologies of \eagle{} galaxies are a close match to those of real galaxies,
at least at low redshift. However, some discrepancies do exist. Mainly, the
distribution of \GAMA{} galaxies appears to be skewed towards higher $G$ and
more negative M$_{20}$ values, as compared to Ref-100 galaxies. This results in
a larger proportion of real galaxies in the E/S0/Sa sector of the morphological
space. Some of this discrepancy can be attributed to the higher median stellar
mass of \GAMA{} galaxies, as discussed in sections
\ref{sec:the_observational_galaxy_sample}.

The discrepancies are much more pronounced for \illustris{}, for which the whole
distribution is skewed towards lower $G$ and more positive M$_{20}$ values,
forming an extended tail up to $M_{20} \simeq -0.5$ where almost no
observational counterparts can be found. These discrepancies are more notable
when considering that all three compared samples are very similar in stellar
mass. 

Recently, \citet{rodriguez-gomez_optical_2019} studied the optical
non-parametric morphologies of galaxies in \illustrisTNG{}, they found that the
updated Illustris model produces galaxies with morphologies much closer to
observations. Indeed, we find that the locus of their $G$-M$_{20}$ distribution
is close to what we find for Ref-100. It is interesting that both simulations,
run with different physical models appear to result in very similar
morphologies. As a matter of fact, there is a better agreement in the
distribution of $G$ and M$_{20}$ between Ref-100 and \illustrisTNG{} than
between any of the simulations and the \GAMA{} galaxies. A possible explanation
for this is that in simulations, the stellar component is represented by
particles tracing the stellar density distribution, and as such, particle noise
gives a granular appearance to the images even when a significant smoothing is
applied. This could explain the shift towards higher M$_{20 }$ values in the
simulations, compared to \GAMA{}. Also, the gravitational softening adopted in
the simulations affects the distribution of matter at the nucleus of galaxies,
resulting in an artificial flattening of the central surface brightness that
could skew $G$ values lower.

Figure \ref{fig:G_M20_images} shows gri-composite images of representative
Ref-100 galaxies at different points in the $G$-M$_{20}$ plane. To construct
the figure we bin galaxies by their $G$-M$_{20}$ values and display the image
of the galaxy closest to the mean value of the bin. Visual inspection reveals
that giant ellipticals and Sa type morphologies dominate the upper right sector
of the figure. While galaxies with more prominent spiral arms (Sb--Sc types) are
more frequent in the lowermost sector. Some small and roundish systems can also
be found in this sector, especially close to the central part of the diagram
($G$ $\sim 0.52$, M$_{20} \sim -1.75$). 

Some of the galaxies found above the demarcation line separating mergers from
normal galaxies show signs of disturbance. However, a majority appears to be
normal, with a morphology not much different to that of galaxies located below
the line. Previously, \citet{bignone_non-parametric_2017} found that
\illustris{} galaxies in this region of the $G$-M$_{20}$ space could be
associated to recent and ongoing mergers, with some contamination from normal
galaxies. It is possible that the demarcation line between mergers and normal
galaxies be shifted in \eagle{} or that the merger properties in the simulation
differ from observations. Recently, \citet{pearson_identifying_2019} tested
whether a convolutional neural network trained on SDSS data could be used to
identify mergers in \eagle{}. They found that the network performed
significantly worst when applied to the simulation, possible indicating
differences between the visually selected observed mergers and the mergers
selected in the simulation.

\subsubsection{Bulge statistic}

Similarly to \citet{snyder_galaxy_2015}, we define a quantity which is a measure
of the optical bulge strength. Specifically, $F$ is defined as five times the
point-line distance from the galaxy's morphology point to the
\citet{lotz_galaxy_2008} early/late type separation line. We also set the sign
of $F$ so that positive (negative) values indicate bulge-dominated
(disc-dominated) galaxies. 

\begin{equation}
    \begin{aligned}
        |F| &= -0.693 * M_{20} + 4.95 \, \textnormal{G} - 3.96, \\
        F(G, M_{20}) &= \begin{cases}
                            |F| \qquad G \geq 0.14 * M_{20} + 0.80, \\
                            -|F| \qquad G < 0.14 * M_{20} + 0.80. \\
                        \end{cases}
    \end{aligned}
\end{equation}

\begin{figure}
    \includegraphics[width=\columnwidth]{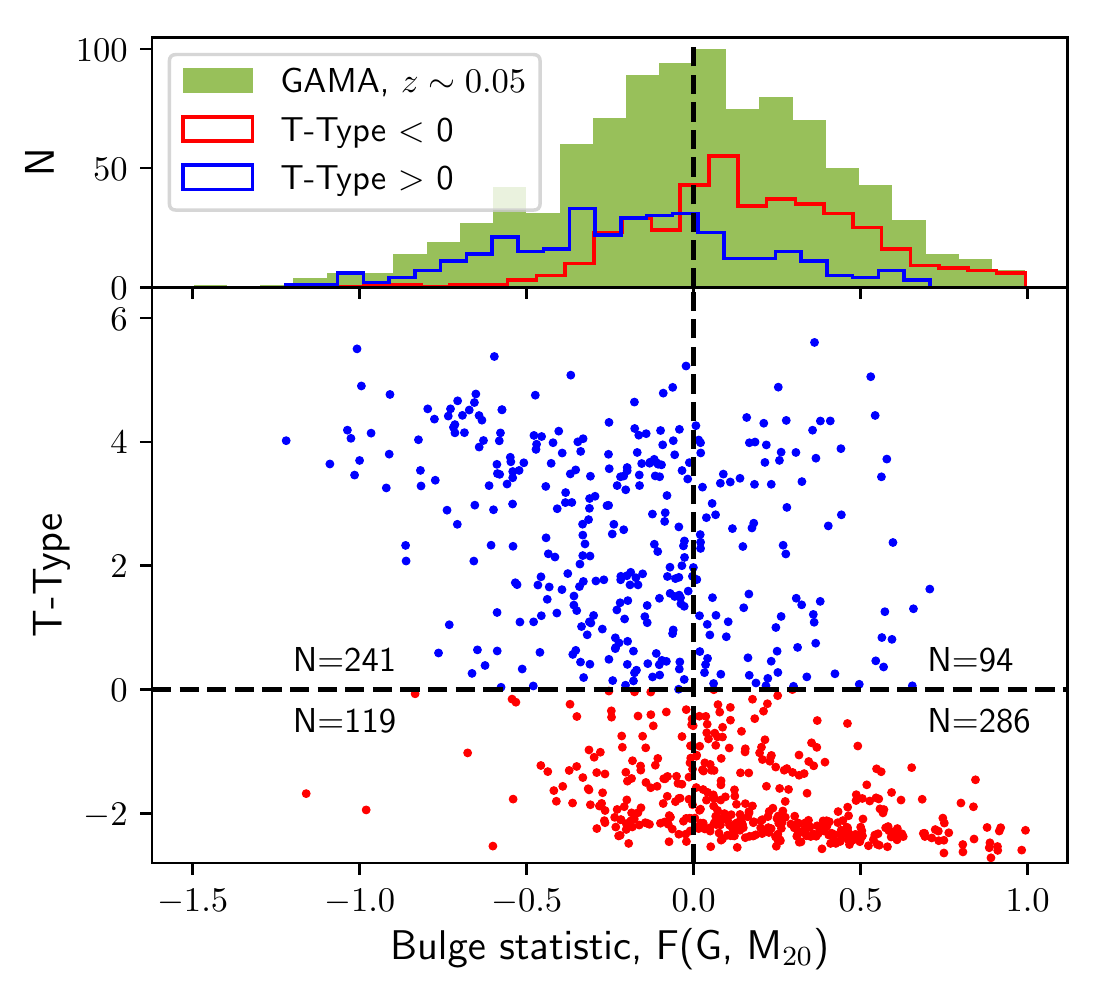}
    
    \caption{T-type versus $F$ for galaxies in the \GAMA{} sample where T-types were assigned
    by a deep convolutional neural network trained in visually classified
    galaxies \citep{dominguez_sanchez_improving_2018}. Red points represent
    negative T-type (bulge-dominated) objects, while blue points represent
    positive T-type (disc-dominated) objects. The bulge strength indicator $F$
    is mostly successful at separating the early and late types as shown by the
    normalized $F$ distributions at the top panel. The top right and bottom left
    corners of the figure contain the minority of objects for which the T-type
    and $F$ classification are in conflict (see text for more details).
    }
    
    \label{fig:F_gama_dist}
\end{figure}

Figure \ref{fig:F_gama_dist} shows the distribution of $F$ for \GAMA{} galaxies,
differentiating positive and negative T-Type populations. We can appreciate that
the $F=0$ separation line is located very close to the point where the number of
early type galaxies starts to dominate. We can also ascertain the level of
contamination that using only $F$ as an assessment of morphological type would
entail. A total of 94 galaxies (28 per cent of T-Type $>$ 0 galaxies) are
classified as late-type according to their T-Type, but as bulge-dominated
according to $F$. On visual inspection, a large number of these systems appear
to be edge-on discs or low-contrast discs which the machine learning algorithm
is able to classify, but that represent a challenge for simple heuristics
derived from non-parametric statistics. 

\begin{figure*}
    \includegraphics[width=\textwidth]{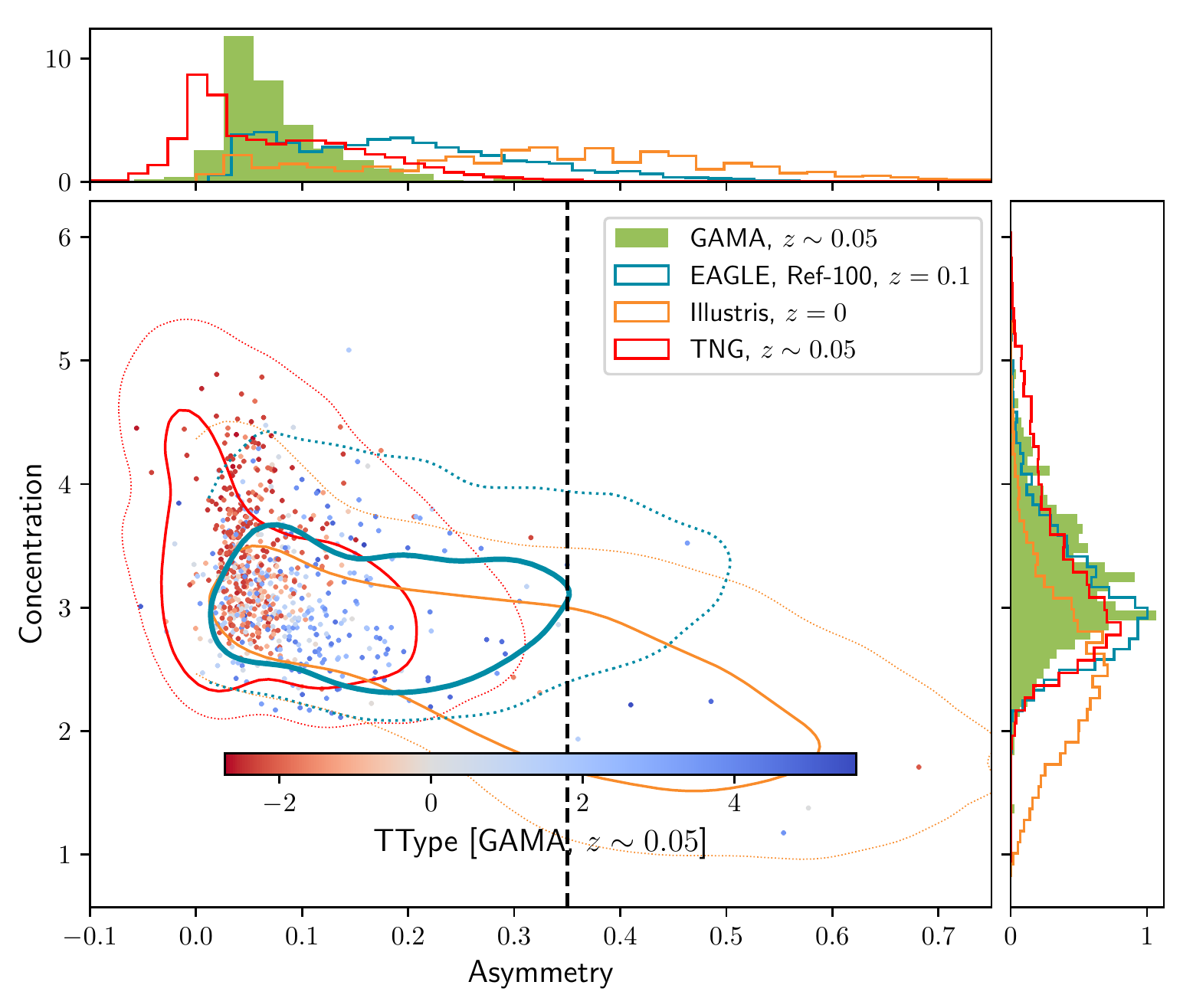}
    
    \caption{The central panel shows the $C$-$A$ diagram from
        galaxies in Ref-100 (blue), \illustris{} (orange), \illustrisTNG{}
        (red) and \GAMA{} (points). The coloured Solid (dotted) lines enclose
        regions containing 68 (95) percent of galaxies in each respective
        sample. The \GAMA{} galaxies are coloured according to their T-Types.
        The black dashed line at $A$=0.35 separates normal from merging or
        highly disturbed galaxies. The distribution of $C$s for
        Ref-100 and \illustrisTNG{} are in good agreement with that of \GAMA{},
        while \illustris{} exhibits a tail towards lower lower $C$s.
        All simulations have a tail towards higher asymmetries in excess of what
        is observed, the effect is more notorious for \illustris{}. $A$
        for \illustrisTNG{} galaxies appear systematically shifted towards
        lower values, this is due to slight changes in the algorithm used to
        compute the statistic, see text for details.}
    
    \label{fig:asymmetry_concentration_scatter}
\end{figure*}

\begin{figure*}
    \includegraphics[width=\textwidth]{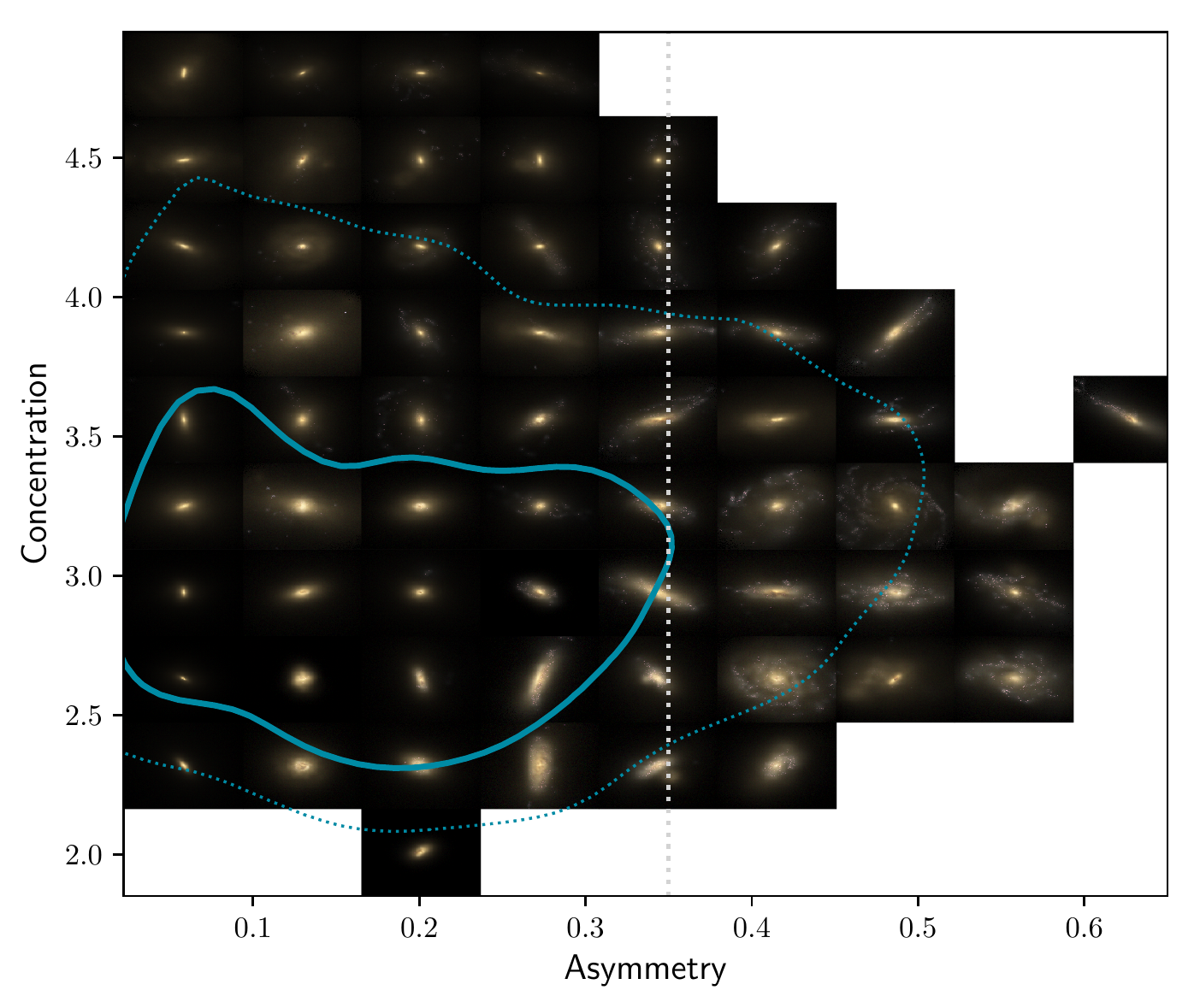}
    
    \caption{Mock gri-SDSS colour composite images of galaxies in the Ref-100
    sample arranged according to their $C$ and $A$ values. Solid
    (dotted) contours represent the region containing 68 (95) percent of
    objects. The straight line is as in Figure
    \ref{fig:asymmetry_concentration_scatter}. In accordance with observational
    trends, prevailing morphologies at the top left of the diagram are of the
    early type, while galaxies at the lower right are late-type.}
    
    \label{fig:A_C_images}
\end{figure*}

The other source of conflict comes from galaxies classified as early types by
their T-Type, but as disc-dominated by $F$. There are 119 cases of this (29 \%
of T-Type $<$ 0 galaxies). Their location in Fig. \ref{fig:F_gama_dist}
indicates that they belong to the same grouping as $F>0$ galaxies, indeed visual
inspection reveals an abundant number of S0 type galaxies.
\citet{dominguez_sanchez_improving_2018} discuss the difficulty of their
algorithm in differentiating between pure ellipticals and S0s, with the
elliptical classification being preferred due to the larger number of training
examples. This suggests that a T-Type closer to zero would actually be a better
match to the morphology of these conflicting galaxies. This will also result in a
tightening of the correlation between T-Type and $F$ that appears in Fig.
\ref{fig:F_gama_dist}. 

These results confirm the robustness of the demarcation line to separate late
and early-type morphologies. We find there is no clear alternative demarcation
line in $F$ to that of \citet{lotz_galaxy_2008} that better separates positive
and negative T-Types.

\subsection{Concentration-Asymmetry}

Figure \ref{fig:asymmetry_concentration_scatter} shows the position in the
$C$-$A$ morphological subspace occupied by the Ref-100 sample
(blue contours), the \illustris{} sample (orange contours), the \illustrisTNG{}
sample (red) and the \GAMA{} sample (coloured points). The subspace is divided into
two sectors by a vertical line at $A=0.35$ which serves to separate mergers from
normal galaxies \citep{lotz_evolution_2008}. The observational \GAMA{} sample
exhibit the expected trends between T-type morphology and non-parametric
statistics with giant ellipticals presenting high $C$ and low
$A$ and late-type disks (Sc–Sd) presenting low $C$ and high
$A$. Intermediate cases appear mixed at approximately $C \sim 3$, $A \sim
0.07$. 

It is clear from Fig. \ref{fig:asymmetry_concentration_scatter} that while the
$C$ distribution of the Ref-100 and \GAMA{} samples are a close match,
that is not the case of the $A$ distribution. Simulated galaxies exhibit a
large tail towards higher asymmetries that do not match observations. Similar
results were obtained for the \illustris{} simulation
\citep{bignone_non-parametric_2017}. Simulated asymmetries have a bimodal
distribution with a low $A$ population that approximately follows
observational trends and another, highly asymmetrical population, for which
there are no observational counterparts. 

For \illustrisTNG{} we find a similar behaviour as Ref-100, but with $A$
shifted towards lower values. This is largely a consequence of the different
implementation of the computation of asymmetries between the simulated samples,
as described in Section \ref{sec:code_asymmetry}. \illustrisTNG{} also shows a
larger tail towards high $C$ galaxies, compared to Ref-100. These
galaxies also correspond to systems with higher $G$ coefficients and M$_{20}$,
compared to Ref-100 and primarily affects massive galaxies. 

Figure \ref{fig:A_C_images} shows colour-composite of Ref-100 galaxies arranged
by their position in the $C$-$A$ plane. Normal spiral galaxies
are mostly found with asymmetries well beyond 0.35, which normally would
indicate disturbed morphologies. It is clear that asymmetry is being driven by
the light distribution of young star-forming regions in the simulated galaxies.
In fact, very young star formation regions are conspicuous in every image where
they are present, because of their scattered and point-like appearance. This is
true even for galaxies with a bulge-dominated morphology. It is therefore likely
that this is an effect of the way the simulated data is being translated into
the mock images for this young stellar populations, specifically the way photon
sources are spatially distributed \citep{torrey_synthetic_2015}. A possible
mitigation strategy could be to assign young stellar particles an increased
smoothing length in the mock image generation procedure
\citep{trayford_optical_2017}.

\subsection{Spatial resolution Effects}
\label{sec:resolution_effects}

\begin{figure}
    \includegraphics[width=1\columnwidth]{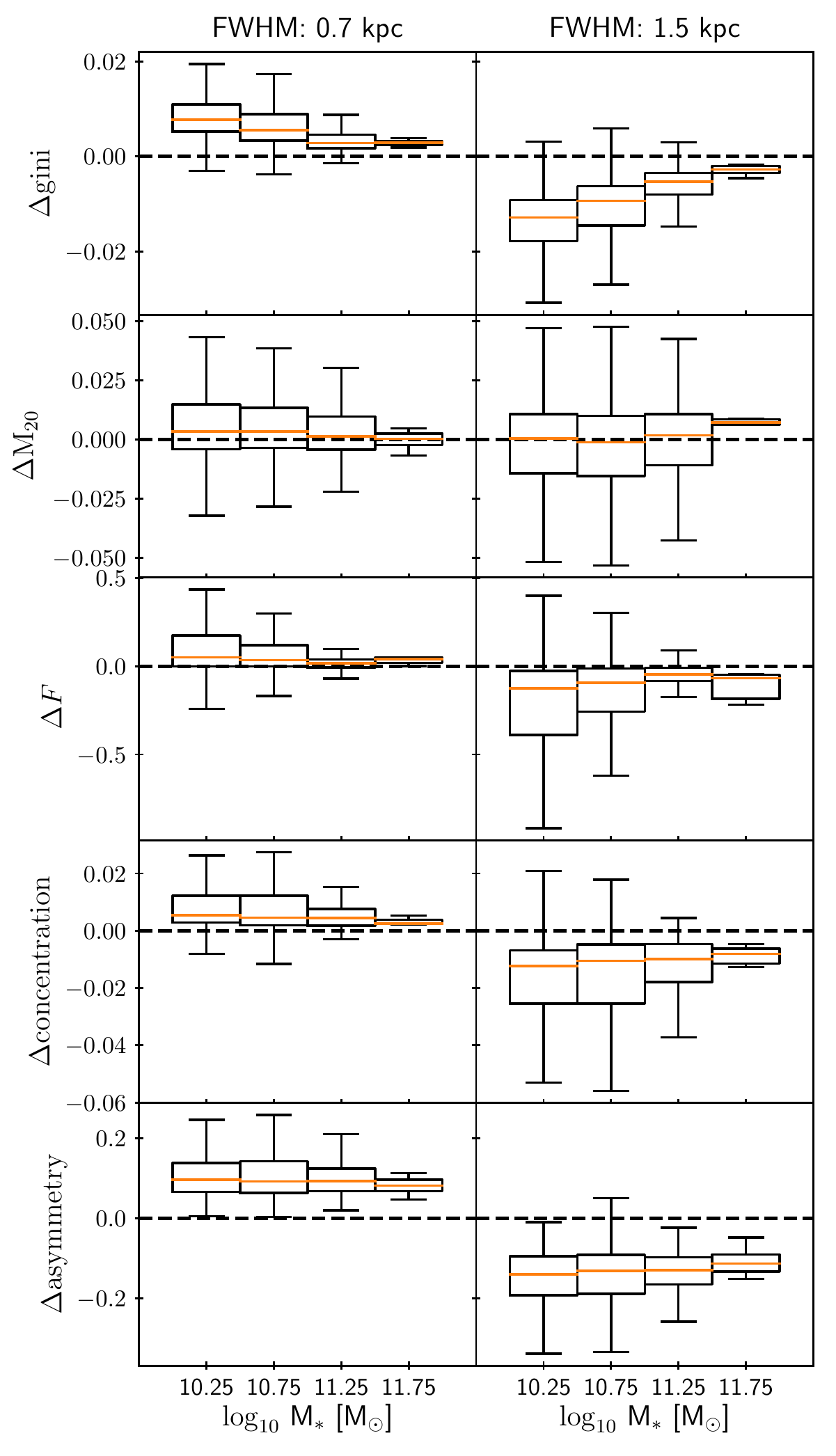}
    
    \caption{Boxplots describing the median relative changes in morphological
        values obtained as a consequence of varying the spatial resolution of
        images by using different FWHM values in the procedure described in
        Section \ref{sec:realistic_images}. Changes in the statistics are
        measured from those obtained using a FWHM=1.0 kpc. $G$ values are
        systematically reduced with decreasing resolution, but the effect is
        larger for decreasing stellar mass. $F$ is the most robust statistic in
        terms of changes in the spatial resolution, while $A$ is the most
        affected overall.}
    
    \label{fig:resolution_effects}
\end{figure}

Non-parametric morphologies can be affected by several factors. Among them,
limited image spatial resolution. Understanding these effects is important,
particularly when contrasting local against high-redshift galaxies, where the
signal-to-noise ratio and the spatial resolution are expected to be worse. Also,
large-scale galaxy surveys (such as SDSS and LSST) which are ideally suited for
statistical studies because of their large sample sizes and the comprehensive
sets of measured quantities, likely suffer from limited resolution.

Previously, \citet{lotz_new_2004} studied the effect of decreasing spatial
resolution on the values of $G$, M$_{20}$, $C$, and $A$. They
found that $C$ and M$_{20}$ were reliable up to resolution scales of
500 pc pixel$^{-1}$, while $G$ and $A$ where stable down to 1000 pc
pixel$^{-1}$. However, their results were restricted to a small sample of 8
galaxies of various morphological type. Here, we have the advantage of a much
larger number of simulated galaxies that also happen to cover a wide range of
morphologies, stellar masses, star formation rates (SFRs) and orientations.
Therefore, we can give a more statistically reliable assessment of the effect of
spatial resolution on non-parametric morphologies.

To study the effect of decreasing resolution we vary the value of the FWHM used
to approximate the seeing in the procedure described in section
\ref{sec:realistic_images}. We consider FWHM values equal to 0.7 kpc, 1.0 kpc
and 1.5 kpc. Results for the intermediate FWHM $= 1.0$ kpc are shown thought
this work and constitutes our value of reference. At 0.7 kpc, the first value
represents an instrument with the same spatial resolution as in the Ref-100
simulation. Also, for $z \sim0.05$ galaxies a FWHM = 0.7 kpc produces images
with the expected spatial resolution of the upcoming LSST, which will have a
mean seeing of 0.7 arcsec. Finally, the last FWHM value more closely represent
the seeing present in SDSS.

\begin{figure*}
    \includegraphics[width=1\textwidth]{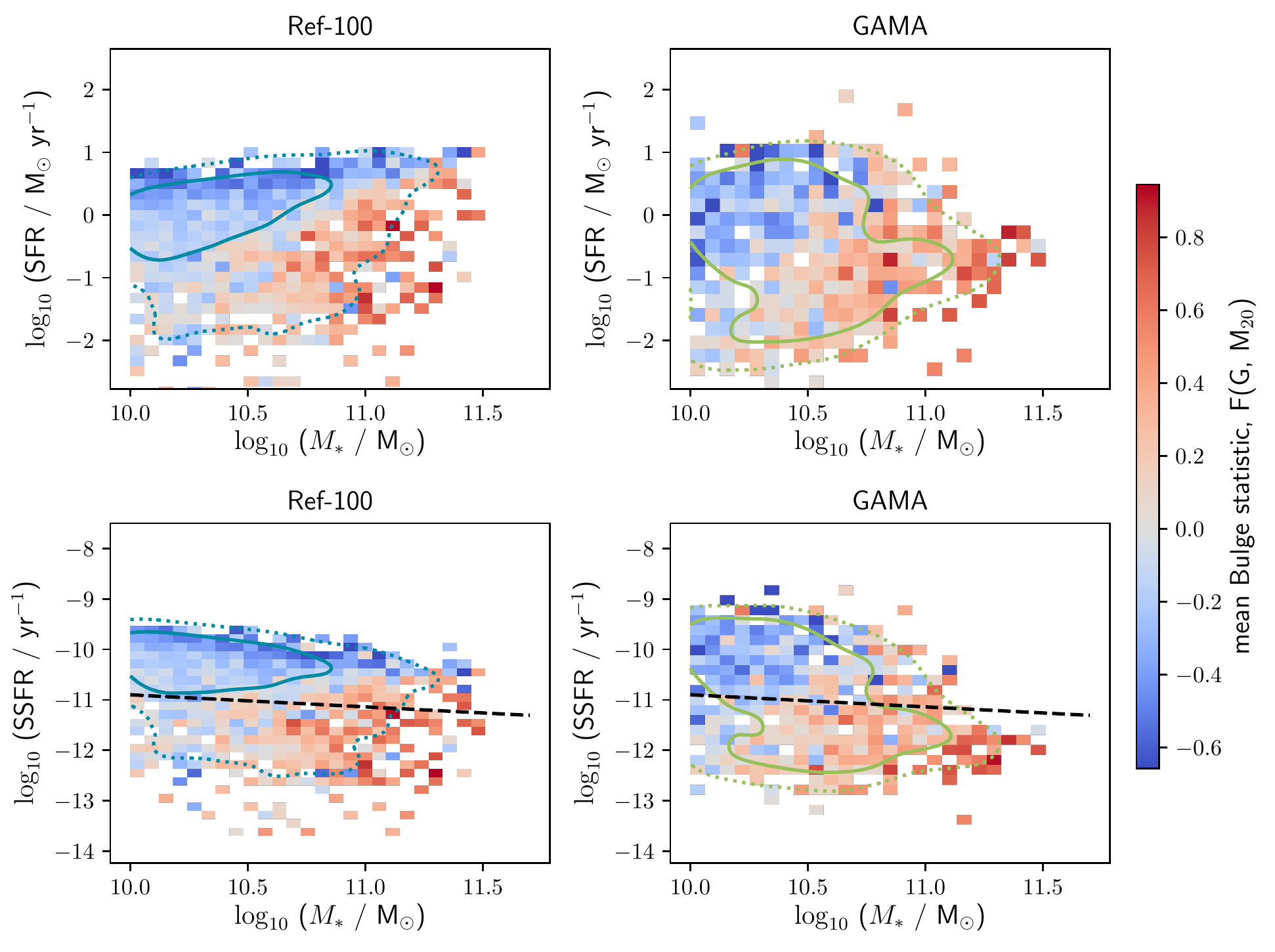}
    
    \caption{Top panels: SFR versus stellar mass for galaxies in the Ref-100
    (left) and \GAMA{} samples (right). with colours proportional to the mean
    bulge strength $F$ in each 2D bin. Solid (dotted) lines enclose regions that
    contain 68 (95) percent of objects. Bottom panels: the same as the top
    panels but for SSFR versus stellar mass. The dashed line separates active
    from passive galaxies according to the criteria of
    \citet{omand_connection_2014}. Ref-100 presents very similar trends in
    stellar mass, star formation and optical morphology compared to \GAMA{}, the
    most notable difference being an excess of active galaxies in the upper
    stellar mass end of the relation for Ref-100, for which there are no
    observation counterparts.}
    
    \label{fig:sfr_dist}
\end{figure*}

We divided the Ref-100 galaxy sample into four subsamples according to stellar
mass. In Fig \ref{fig:resolution_effects} we show boxplots describing the median
relative changes in morphological values obtained as a consequence of using
different FWHM values for each subsample. Changes are measured from results
obtained with FWHM $= 1.0$ kpc according to:
\begin{equation}
    \Delta_{X_i} = \frac{X_i - X_{1.0 \,\textnormal{kpc}}}{\left|X_{1.0 \,\textnormal{kpc}}\right|},
\end{equation}
where $X$ represents $G$, $M_{20}$, $F$, $C$ or $A$, while the suffix $i$ stands
for one of the tested FWHM values: 0.7 and 1.5 kpc.

We find that $G$ is systematically reduced with decreasing spatial resolution.
Also, the effect is larger for lower mass galaxies. For FWHM=0.7 (1.5) kpc and
stellar masses $\sim 10^{10.25}$ \MsunInline{}, the median change in $G$ with
respect to the reference values is $\sim 0.8$ ($\sim 1.5$) per cent higher
(lower). In contrast, M$_{20}$ is less effected, with median shifts less than
0.8 per cent for both FWHM values, across all mass bins. $F$ is also
systematically reduced with decreasing spatial resolution, this is most
noticeable for FWHM=1.5 kpc, where median shifts in $F$ are $\sim 10$ per cent
towards lower values. Also, there is a large scatter in $\Delta_{F_i}$,
specially in the two lower mass bins. Results indicate that G-M$_{20}$ values of
larger mass galaxies are comparably more robust. This can also be appreciated in
Figure \ref{fig:gini_m20_scatter} where both simulated and observed galaxies
with $F < 0$ appear to move away from the \citet{lotz_evolution_2008} line in
the direction predicted by the resolution effect. $F > 0$ galaxies on the other
hand, have a distribution parallel to the \citet{lotz_evolution_2008} line. This
behaviour can be easily explained by the smoother light distribution of
early-type galaxies that is largely unaffected by additional smoothing by the
seeing. These spatial resolution effects could also explain why the demarcation
line was found to be slightly different between the \citet{lotz_new_2004} and
\citet{lotz_evolution_2008} studies, since the latter study was based on a
closer sample of galaxies with the consequential higher spatial resolution.

$C$ only shows a small systematic effect of less than 2 per cent even for the
1.5 kpc worst case scenario, and a small dependence on stellar mass. While the
quantity most affected by spatial resolution is $A$, which exhibits values 14
per cent lower for FWHM=1.5 kpc. However, simulated asymmetries are considerably
larger than observed ones, as previously discuss, so it is likely that this
effect is a product of the simulated nature of the images and not directly
applicable to observational results.

\subsection{Dependence on star formation}
\label{sec:dependence_on_star_formation}

Measurements of S\'ersic index and compactness are found to correlate with
galaxy quiescence \citep[e.g.][]{wuyts_galaxy_2011, bell_what_2012} indicating
that galaxy morphology and star formation are closely related. 

In Figure \ref{fig:sfr_dist} we plot the mean values of the bulge statistic $F$
in bins of (SFR, M$_*$) and (SSFR, M$_*$). To each of these mean $F$ values we
assign colours from blue (disc-dominated) to red (bulge-dominated). We also plot
contours containing 68 percent (solid lines) and 95 percent (dotted line) of the
galaxies in each sample. The star formation rate is extracted directly from the
simulation in the case of Ref-100 and from H$\alpha$ luminosity measurements for
\GAMA{} galaxies \citep{gunawardhana_galaxy_2013}.

We find that Ref-100 galaxies roughly recover the main sequence of star-forming
galaxies \citep[e.g.][]{whitaker_star_2012}. Although, results by
\citet{furlong_evolution_2015} showed that the Ref-100 simulation presented
SSFRs $\sim0.2$ dex lower compared to other observational data sets
\citep{gilbank_local_2010,bauer_galaxy_2013}. Despite these possible offsets in
the normalization, we find that in general terms lower SFR galaxies of the same
stellar mass have, on average, a more bulge-dominated morphology. There is a
good agreement between the SFR-M$_*$-F relation of simulated and observed
samples. 

We also find that the bulge-dominated morphologies are mostly found for stellar
masses $> 10^{10.5}$ \MsunInline{} and with SSFRs consistent with quenched star
formations, as indicated by their position below the line separating active and
passive galaxies according to \citet{omand_connection_2014}. However, some
bulge-dominated systems can still be found among star-forming galaxies
\citep{rosito_field_2018}. Also passive galaxies can have disc morphologies, but
these tend to be relegated to lower mass systems. 

There is a population of star-forming and high-mass galaxies in Ref-100 for
which there is no equivalent among the \GAMA{} sample. This suggests that the
quenching mechanisms in the simulation are not efficient enough in these
particular cases. This is in line with results by \citet{furlong_evolution_2015}
who found $\sim 15$ per cent too few passive galaxies between $10^{10.5}$ and
$10^{11.5}$ \MsunInline{} in Ref-100, compared to observations. We find that the
Morphologies of these galaxies are mostly late-type, but early-types start to
dominate at lower SFRs.

\subsection{Dependence on size}

\begin{figure*}
    \includegraphics[width=1\textwidth]{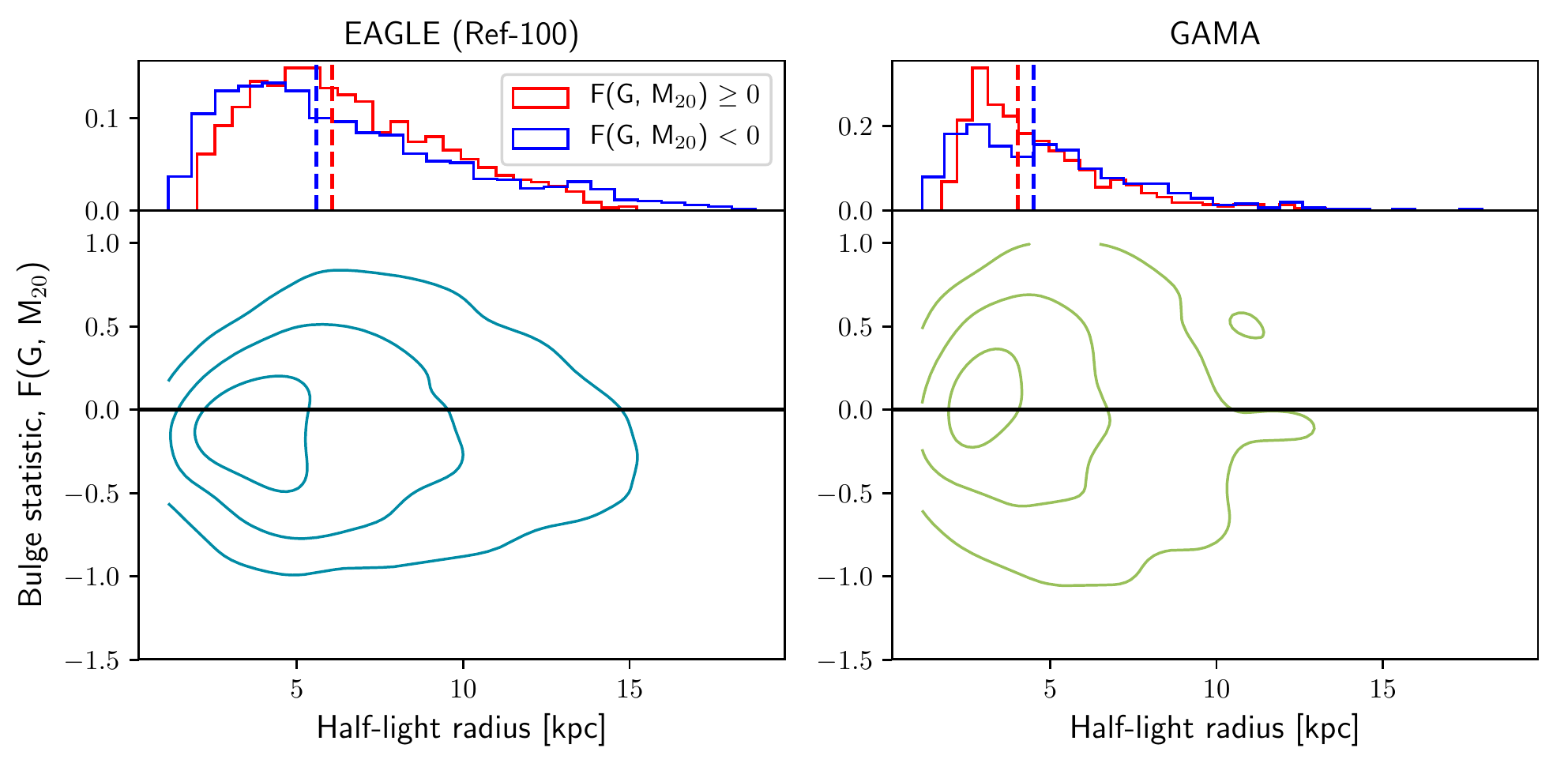}
    
    \caption{Bulge strength statistic $F$ versus galaxy size parametrized by the
        semimajor axis of an ellipse containing half of the total flux. The
        panel on the left shows galaxies in Ref-100, while the panel on the
        right shows galaxies from \GAMA{}. The contours indicate the overall
        distribution of galaxies, while the histograms at the top panels
        indicate the normalized distribution of galaxy sizes discriminating
        between early ($F > 0$) and late ($F < 0$) optical morphologies. The
        coloured dashed lines in the histograms represent the median half-light
        radius of each subsample. There is an approximate agreement between
        Ref-100 and \GAMA{} in terms of the optical morphology dependence on
        size. Although the correlation is slightly stronger for \GAMA{}
        galaxies, meaning that \GAMA{} galaxies with more disc-dominated
        morphologies present slightly higher median sizes.}
    
    \label{fig:F_half_light}
\end{figure*}

The bottom panels of Figure \ref{fig:F_half_light} show the bulge statistic $F$
as a function of galaxy size. The galaxy size is parametrized by the semimajor
axis of an ellipse containing half of the total luminosity of the galaxy. Upper
panels show the size distribution of galaxies discriminating between
bulge-dominated ($F \geq 0$) and disc-dominated ($F < 0$) systems. Both Ref-100
and \GAMA{} samples present similar flat distributions, with the observational
data presenting a slightly higher degree of correlation between disc strength
and galaxy size. Meaning that \GAMA{} galaxies with more disc-dominated
morphologies present slightly higher median sizes.

\citet{furlong_size_2017} studied the evolution of galaxy sizes in the \eagle{}
simulations. They found that the dependence of the sizes of simulated galaxies
on stellar mass and star formation is close to that of observed galaxies. They
also found that active galaxies are typically larger than their passive
counterparts at a given stellar mass. This is in general agreement with the
results we find for the \GAMA{} sample.

\begin{figure*}
    \includegraphics[width=\textwidth]{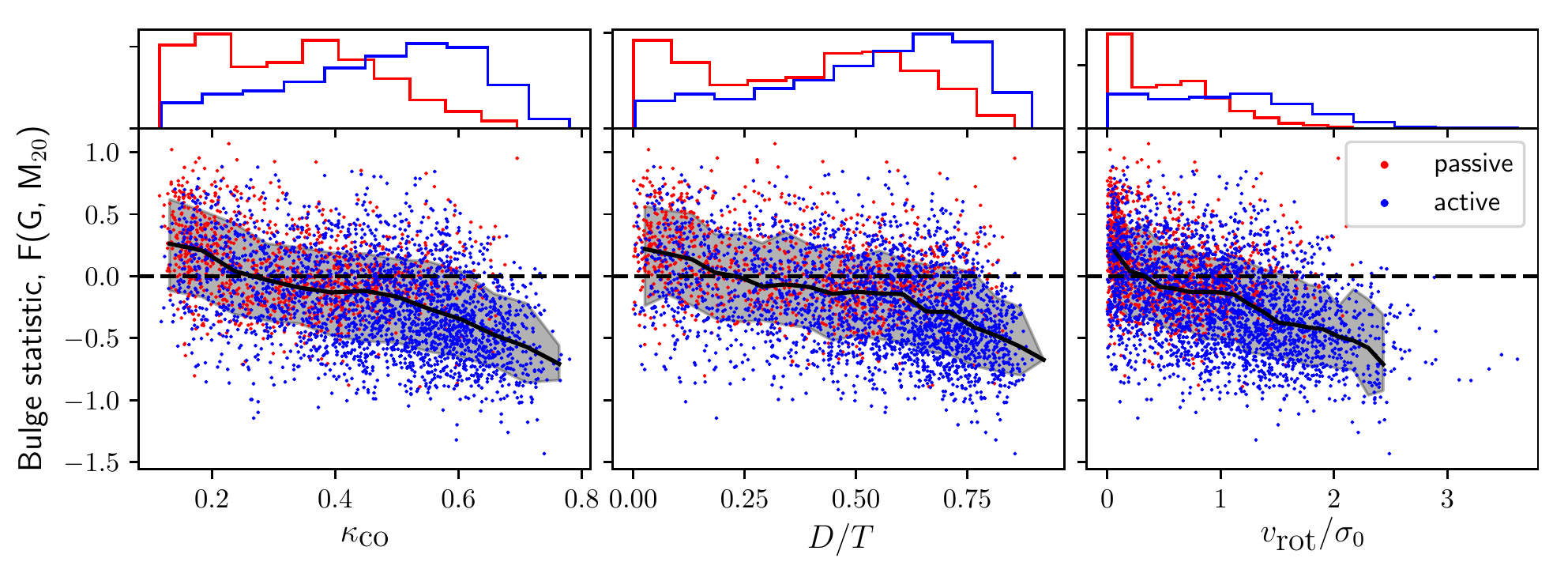}
    
    \caption{The bottom panels show the bulge strength statistic $F$ versus
    $\kappa_\textnormal{co}$ (left), D/T (centre) and
    $v_\textnormal{rot}/\sigma$ (right) for Ref-100 galaxies. The solid lines
    show the binned median and 1$\sigma$ (16th-84th) percentile scatter of the
    dependent variables. Overall, the optical morphology shows a strong
    anti-correlation with the kinematic metrics of morphology. The bottom panels
    show the normalized distribution of each kinematic metric for active (blue)
    and passive (red) galaxies.}
    
    \label{fig:kinematics_F}
\end{figure*}

Similar comparisons between morphology and galaxy size are discussed in
\citet{rodriguez-gomez_optical_2019} for the cases of \illustris{} and \illustrisTNG{}. They
found that, while late-type \illustris{} galaxies are indeed larger than their
early-type counterparts, the inverse is true for \illustrisTNG{} galaxies. Meaning that
there is tension in the size-morphology relation between observed and \illustrisTNG{}
galaxies. 

It should also be pointed out that \illustris{} galaxies are about two times
larger than observations at $z=0$ and that \illustrisTNG{} galaxies show an overall better
agreement with observations in terms of sizes and observational qualitative
trends of size with stellar mass, star formation rate and redshift
\citep{genel_size_2017}. In general terms, this means that both \illustris{}
simulations are in tension with observations, albeit for different reasons.
Ref-100 galaxies, however, show no significant tension with the GAMA results as
shown in \citet{furlong_size_2017} and by the present results.

\subsection{Dependence on rotation}

\begin{figure*}
    \includegraphics[width=\textwidth]{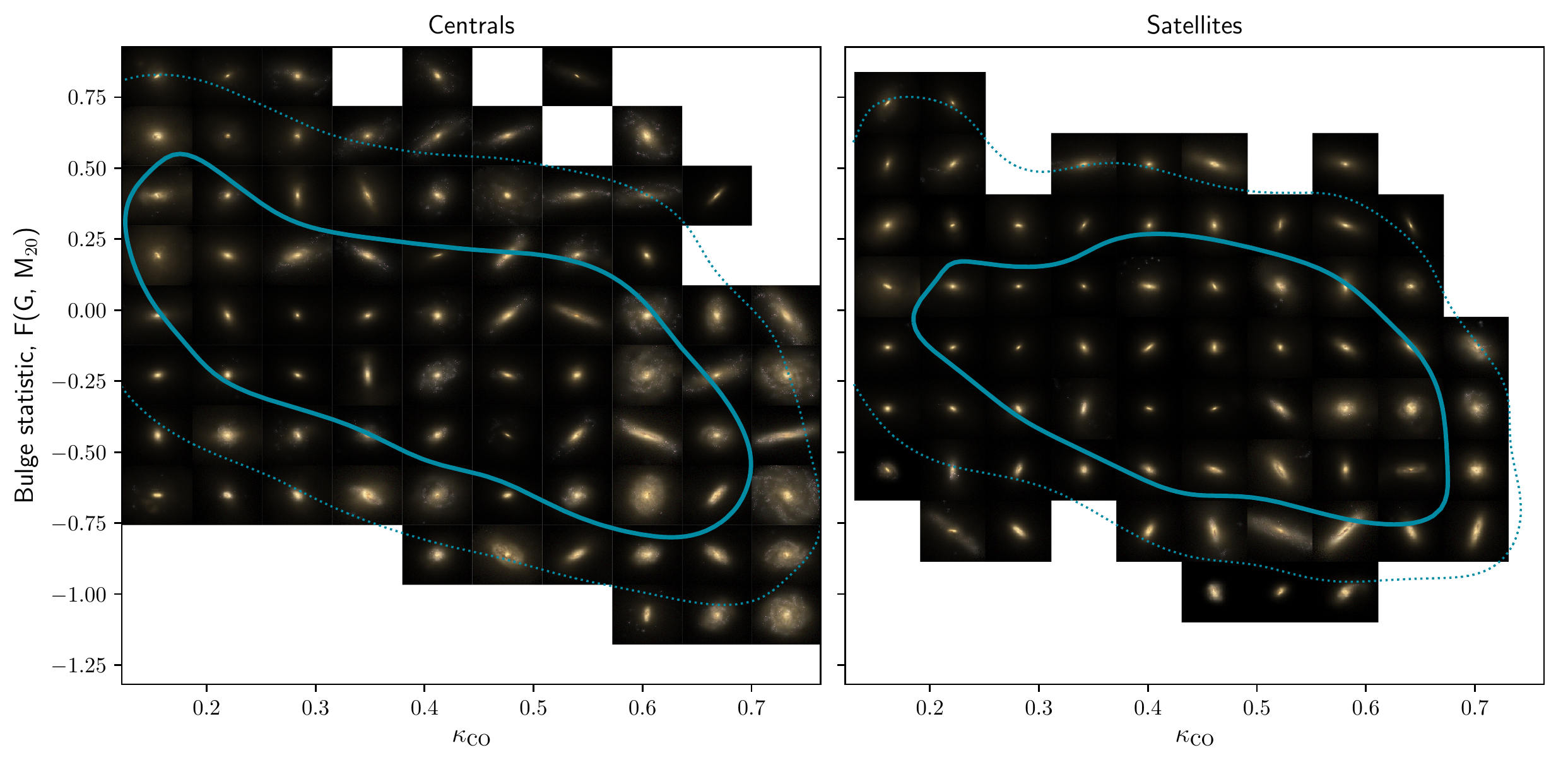}
    
    \caption{Mock gri-SDSS colour composite images of galaxies in the Ref-100
        sample arranged according to their $F$ and $\kappa_\textnormal{co}$
        values. The left (right) panel contains central (satellite) galaxies.
        Solid (dotted) contours represent the region containing 68 (95) percent
        of objects. The Figure illustrates the general trend of bulge-dominated
        galaxies appearing on the top left of the diagram while disc-dominated
        appear mostly on the lower right corner. Central galaxies with
        $\kappa_\textnormal{co}\sim0.5$ but high $F$ values exhibit a disky
        appearance but with a prominent bulge. Satellites have a distinct
        appearance from their central counterpart at equal ($F$,
        $\kappa_\textnormal{co}$) values. In general, they appear smoother and
        with less prominent discs.}
    
    \label{fig:kinematics_imgs}
\end{figure*}

\begin{figure*}
    \includegraphics[width=\textwidth]{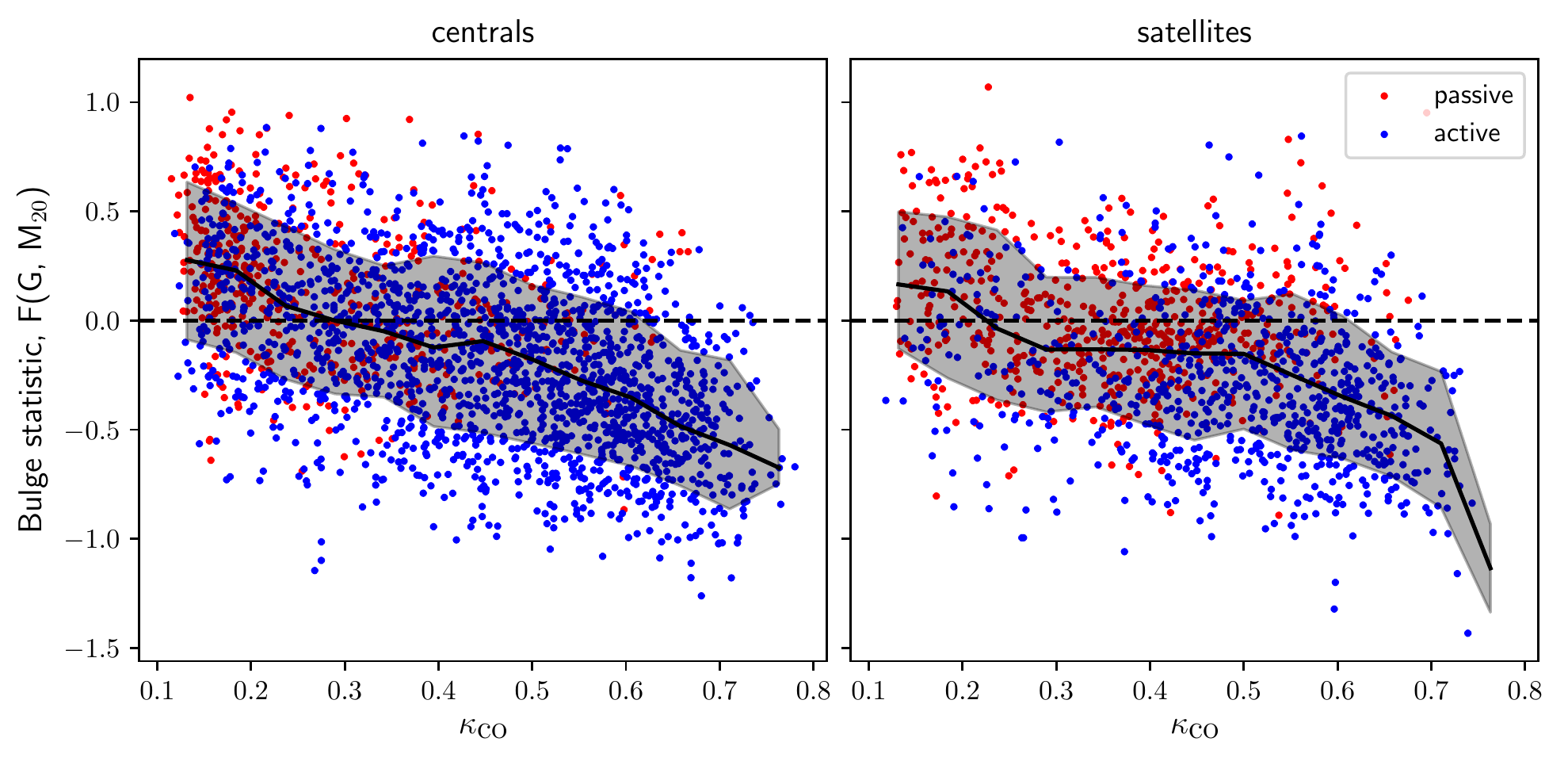}
    
    \caption{As the first panel of Figure \ref{fig:kinematics_F} but
        discriminating between central (left panel) and satellite (right panel)
        galaxies. Satellites show less correlation between $F$ and
        $\kappa_\textnormal{co}$ than centrals, this is mainly due to a
        population of lower stellar mass, quenched satellites that occupy the
        central location in the relation. These galaxies present a similar
        bulge-dominated appearance independently of their rotational support.}
    
    \label{fig:kinematics_F_centrals_sats}
\end{figure*}

There is a clear correlation between the internal kinematics of galaxies and
their morphological appearance. In general terms, disc galaxies have been shown
to be supported by rotation, while spheroidal systems, such as ellipticals are
supported by dispersion. However, recent surveys have revealed that the
connection between internal kinematics and morphology is not straightforward. In
particular, the stellar angular momentum of early type galaxies have been found
to span a range of values from slow to fast rotators, while a majority of S0
galaxies have been found to be fast rotators \citep{emsellem_atlas3d_2011-1},
suggesting that early and lenticular types belong to the same class but
differentiate on their degree of rotational support. This indicates that
kinematic diagnostics might give a more fundamental and physically motivated
classification scheme \citep[e.g.][]{emsellem_sauron_2007,
krajnovic_sauron_2008, cappellari_atlas3d_2011}. Strong correlations between
optical morphology and rotation have also been found for late-type galaxies,
suggesting the existence of a fundamental relation between angular momentum,
stellar mass and optical morphology across all Hubble types
\citep[e.g][]{romanowsky_angular_2012, obreschkow_fundamental_2014, cortese_sami_2016}

Since stellar and gas kinematics are easily extracted from simulations,
kinematic diagnostics have long been used as a proxy for optical morphology.
These diagnostics generally summarize galaxy kinematics in a single parameter
such as the $\kappa_\textnormal{rot}$ parameter \citep{sales_feedback_2010}, the
bulge-to-total ratio (B/T) or the disc-to-total ratio (D/T)
\citep[e.g.][]{scannapieco_effects_2008}. In the case of \eagle{}, variations
of these metrics have been studied by
\citet{correa_relation_2017,correa_origin_2019}, \citet{clauwens_three_2018},
\citet{trayford_star_2019} and \citet{tissera_oxygen_2019-1}.

In Figure \ref{fig:kinematics_F} we show the optically derived $F$ bulge
strength statistics as a function of three kinematic metrics: $D/T$, the
fraction of kinetic energy that is invested in co-rotation
\citep[$\kappa_\textnormal{{co}}$, ][]{correa_relation_2017} and the ratio of
rotation and dispersion velocities ($v_\textnormal{rot}/{\sigma}$). All three
quantities are extracted from the \eagle{} public database and are based on the
corresponding definitions found in \citet{thob_relationship_2019}. We find that
$F$ anti-correlates with all three kinematic diagnostics to a similar extent, a
Spearman's rank test gives correlation coefficients of -0.46, -0.46 and -0.43
between $F$ and $\kappa_\textnormal{{co}}$, $D/T$ or
$v_\textnormal{rot}/{\sigma}$, respectively. The scatter in $F$ is $~0.7$ dex
for all three kinematic metrics. This shows that the optical morphologies of the
simulated galaxies correlate with the degree of rotational support to a large
extent. The similar correlation coefficients found are in line with results by
\citet{thob_relationship_2019} that show that these commonly used kinematic
metrics are strongly correlated in \eagle{} and can in general be used
interchangeably.

Also in Figure \ref{fig:kinematics_F} we distinguish between active (blue
points) and passive galaxies (red points) using the same criteria as in Section
\ref{sec:dependence_on_star_formation}. It is apparent that
$\kappa_\textnormal{co}$ is the most successful among the kinematic metrics in
separating between star-forming and quenched galaxies as can be appreciated from
the normalized histograms in the top panels, indeed \citet{correa_relation_2017}
showed that simple threshold at $\kappa_\textnormal{co}=0.4$ serves to separate
galaxies in the red sequence from those in the blue cloud. We notice that such a
value of $\kappa_\textnormal{co}$ roughly corresponds to the transition between
optically bulge dominated ($F > 0$) and disc dominated ($F < 0$) galaxies. This
serves to confirm in a quantitative way that that choice of
$\kappa_\textnormal{co}$ threshold is also successful at separating galaxies
that look disky from those that look elliptical. 

However, we also notice that using a threshold in $F$ instead of
$\kappa_\textnormal{co}$ to classify galaxies selects in principle, different
galaxy subsets. In particular, there is a group of active galaxies around
$\kappa_\textnormal{co} \sim 0.5$ that presents positive $F$. These galaxies
would be classified as disc-dominated according to their kinematics, but as
bulge-dominated according to their light distribution. In Figure
\ref{fig:kinematics_imgs} we investigate the visual appearance of galaxies based
on their location in the $F$ versus $\kappa_\textnormal{co}$ space. We confirm
the general trend that early type and late type galaxies are located
respectively in the top left and bottom right of the diagram. We also notice
that the mentioned subset of galaxies with contradicting kinematic and optical
morphologies are mostly central galaxies with active star-forming regions and
tend to have more of a disky morphology. However, they differ from the pure
spirals in that their disc and arms appear less prominent, which would explain
why they are being assigned positive $F$ values. These galaxies could correspond
to disc+bulge galaxies explored by \citet{clauwens_three_2018}.

The right panel of Figure \ref{fig:kinematics_imgs} show galaxy images for
satellite galaxies. Compared to the central galaxies on the left, they present a
somewhat different appearance. For equal ($F$, $\kappa_\textnormal{co}$)
satellites are more compact, present less prominent discs and generally a
smoother appearance, indicating differences in their evolution. This can be
expected if for example. environmental processes result in additional quenching
mechanisms \citep{kauffmann_environmental_2004} in satellites. Figure
\ref{fig:kinematics_F_centrals_sats} further explores the difference between
central and satellite galaxies in the correlation between $F$ and
$\kappa_\textnormal{co}$. We find that optical and kinematic morphology
indicators are more correlated in the case of centrals, indeed a Spearman's rank
test gives correlation coefficients of -0.5 and -0.38 between $F$ and
$\kappa_\textnormal{{co}}$ for central and satellite galaxies, respectively. For
satellites it can be appreciated that there is a flattening in the relation at
$\kappa_\textnormal{co} \sim 0.3$ where there is an abundance of quenched
galaxies. Inspection of Figure \ref{fig:kinematics_imgs} reveals that these are
mostly smallish, lower mass systems that likely experienced environmental
quenching without having gone through a kinematic transformation. This result is
in agreement with \citet{cortese_sami_2019} who found that satellites undergo
little structural change before and during the quenching phase.

In contrast, satellites with low $\kappa_\textnormal{{co}}$ and high $F$ values
appear clustered apart and their visual appearance is more similar to their
central counterparts. This indicates that for these higher stellar mass systems,
environmental quenching is not as important and that their morphological
evolution is more akin to that of central galaxies. 

\section{Summary and discussion}
\label{sec:Summary_and_discussion}

We have studied the optical morphologies of $z=0.1$, $M_* > 10^{10}$
\MsunInline{} galaxies in the \eagle{} Ref-100 simulation using non-parametric
statistics (Gini, M$_{20}$, Concentration and Asymmetry) derived from the g-band
light distribution in mock images obtained from radiative transfer techniques
including the effect of dust and post-processed to mimic images in the SDSS
survey. We have compared the Ref-100 morphologies with those of galaxies from
the \GAMA{} survey and from other numerical simulation, \illustris{} and
\illustrisTNG{}. Morphologies of simulated and observed images were obtained in
a very similar manner using the same \statmorph{} code
\citep{rodriguez-gomez_optical_2019}

Our conclusions can be summarized as follow:

(i) Optical \eagle{} morphologies indicated by their distribution of $G$ and
M$_{20}$ statistics agree well with those derived from SDSS images of \GAMA{}
galaxies selected to have $z\sim0.05$ and $M_* > 10^{10}$ \MsunInline{}, closely
matching the simulated sample selection.

(ii) The ($G$, M$_{20}$) morphologies of \GAMA{} galaxies correlate well with
their T-Type morphologies obtained using a deep neural network trained on visual
classification \citep{dominguez_sanchez_improving_2018}. Moreover, we find that
the demarcation line separating bulge from disc dominated morphologies according
to \citet{lotz_evolution_2008} performs that task very well for our
observational and simulated sample and therefore is a robust basis for the
definition of the bulge strength statistics $F$ \citep{snyder_galaxy_2015}.

(iii) Simulated galaxies from the \illustris{} simulation
\citep{snyder_galaxy_2015} present some discrepancies with those in \eagle{}
and \GAMA{}, particularly for a subset of galaxies at low $G$ and high M$_{20}$
values for which there are no counterparts in the mentioned samples. In
contrast, simulated morphologies of \illustrisTNG{} galaxies
\citet{rodriguez-gomez_optical_2019} agree remarkably well with those in our
Ref100 and \GAMA{} samples. This indicates that there is a convergence between
the simulations in terms of these morphological statistics, possible due to the
fact that both simulations reproduce to a large extent at $z\sim0$ basic galaxy
properties such as stellar mass, size and star formation rate. Given that a
significant portion of galaxy morphology is determined by these factors, this is
perhaps not that surprising. It is still remarkable, that we can make such a
straightforward and direct quantitative comparison between the optical
morphologies of these various simulations and also observations. 

(iv) Although there are disturbed and interacting simulated galaxies present in
the ($G$, M$_{20}$) region commonly assigned to merger and irregular galaxies
\citep{lotz_evolution_2008} we find that there is significant contamination from
normal galaxies. Recently, \citet{pearson_identifying_2019} used a convolutional
neural network to identify mergers in SDSS and in \eagle{} mock images, they
found that the network lost significant performance when trained or applied to
\eagle{} images as compared to SDSS images. This indicates that the visual
appearance of normal and merging Ref-100 galaxies might present discrepancies
when compared to observations.

(v) Further discrepancies are found for the $A$ statistic. Normal Ref100
spiral galaxies have significantly larger asymmetries that their \GAMA{}
counterparts and similar behaviour is observed for \illustris{} and
\illustrisTNG{} galaxies. This is an indication that despite the general good
agreement between observed and simulated morphologies, simulations still present
differences in their more detailed appearance. A likely explanation for this is
that the distribution of photon sources from young stellar population in the
image generation procedure is resulting in artificially high asymmetries. We
suggest that a possible mitigation strategy could be to assign young stellar
particles an increased smoothing length in the mock image generation procedure
\citep{trayford_optical_2017}. Recently, \citet{dickinson_galaxy_2018} presented
the visual morphological classification of \illustris{} galaxies derived from
Galaxy Zoo citizen scientists. They identified significant differences between
\illustris{} and real SDSS images. Specifically, a much larger fraction of
simulated galaxies were classified as presenting visible substructure,relative
to their SDSS counterparts. As per (iii), both \eagle{} and \illustrisTNG{}
appear to better match observations compared to the original \illustris{},
future studies of this kind could determine if these improvements are enough to
also result in a better match with respect to human visual classification. In
that direction we also point out that a similar neural network to the one used
to classify T-type morphologies in (ii) has recently been used to classify
\illustrisTNG{} galaxies \citep{huertas-company_hubble_2019}. The authors found
that the neural network, trained on SDSS visual morphologies was successful at
identifying simulated galaxies in four classes (E, S0/a, Sab and Scd). In
summary, while very detailed morphologies might need further improvements, it
appears that simulations are successful in reproducing general visual
morphologies.

(vi) The large sample size of simulated galaxies spanning a large range of
stellar masses, sizes and morphologies allows us to study in better statistical
detail the effect that spatial resolutions has on the non-parametric
morphologies. This is particularly important in light of future surveys, such as
LSST where this kind of automatic morphological classification is expected to be
implemented on a very large scale \citep{collaboration_lsst_2009}. We vary the
value of the FWHM used to approximate the seeing between 1.0 kpc (the reference
value), 0.7 kpc (the value expected for LSST), and 1.5 kpc (a value that more
closely match SDSS imaging). We find that $G$ is systematically lower for
decreasing resolution and that such effect depends on stellar mass, with the
lower mass galaxies presenting the largest effect. Similar effects are also
found for $F$. $A$ appears to be the statistic most affected by spatial
resolution, with significantly lower $A$ values with decreasing resolution.
Although, no apparent dependence on stellar mass was found, this is in contrast
to what was found by \citet{bignone_non-parametric_2017} in the case of
\illustris{}, for which lower mass galaxies were systematically more asymmetric.

(vii) Ref100 galaxies present the expected trends between optical
morphology (summarized by the bulge strength $F$), stellar mass and star
formation rate, where at equal stellar mass, galaxies with lower SFR have, on
average, a more bulge-dominated morphology. The general trends found for Ref100
and \GAMA{} galaxies are very similar, with some discrepancies. Specifically,
there is an excess of active Ref-100 galaxies at M$_*$ $> 10^{11}$ \MsunInline{}
for which there are no observational counterparts in \GAMA{}, despite the
similar stellar mass distribution of the samples. This is consistent with
previous results by \citet{furlong_evolution_2015} who found $\sim15$ percent
fewer Ref-100 passive galaxies in the stellar mass range $10^{10.5} -
10^{11.5}$. We find that these galaxies present a mix of bulge-dominated and
disc-dominated morphologies.

(viii) There is a general agreement between the size-morphology
relation of galaxies in the Ref-100 and \GAMA{} samples, in that they both
present a similar flat trend in $F$ versus Half-light radius. Although,
disc-dominated \GAMA{} galaxies appear to be slightly larger than their
bulge-dominated counterparts, in line with previous results by
\citet{furlong_size_2017} who found that active galaxies are typically larger
than their passive counterparts at a given stellar mass. 

(ix) We find a significant correlation between the optical morphology of Ref-100
galaxies, characterized by their bulge strength $F$ and kinematic morphologies
expressed by D/T, $\kappa_\textnormal{co}$ and $v_\textnormal{rot}/\sigma$. In
general terms, optically bulge-dominated galaxies have lower rotational support
and higher velocity dispersion.

(x) We find that a threshold value of $\kappa_\textnormal{co}=0.4$
\citep{correa_relation_2017} roughly corresponds to our $F=0$ threshold
separating bulge-, from disc-dominated systems. However, we notice that the
optical and kinematic criteria do not select the same galaxy populations.
In particular, the optical criteria for bulge-dominated systems differs in that
it also selects a number of actively star-forming central galaxies with
significant rotational support and visually disky appearance. These galaxies
differ from pure spirals in that their discs and arms are less prominent and they
present a significant bulge component.  

(xi) Central galaxies present a higher degree of anti-correlation between $F$
and $\kappa_\textnormal{co}$, compared to the one found for satellites. This is
due to satellites presenting a significant number of low stellar mass, quenched
systems with $F\sim0.1$ values and $\kappa_\textnormal{co}$ between [0.3, 0.5].
This is an indication of different morphological evolution in centrals and
satellites, with smaller mass satellites being more affected by environmental
quenching that shuts down star formation leaving the disc initially intact. This
is in line with results by \citet{cortese_sami_2019} who found that changes in
stellar kinematic properties become evident at a later stage and that satellites
tend to remain rotationally dominated. Also, \citet{trayford_it_2016} found that
lower mass Ref-100 galaxies are quenched by environmental effects once they
become satellites. While, higher stellar mass, central galaxies, are quenched
mainly by AGN feedback \citep{trayford_it_2016, bower_dark_2017}. Higher mass
central and satellite galaxies present a similar appearance in terms of both
optical morphology and their degree of rotational support. It is expected that
for higher mass satellites, environmental quenching is not as important and that
their morphological evolution can be more similar to that of central galaxies. 

\section*{ACKNOWLEDGEMENTS}

The authors acknowledge Joop Schaye for useful discussions that helped to guide
this work and Vicente Rodriguez-Gomez for providing the non-parametric
morphologies of \illustrisTNG{} galaxies. The authors acknowledges support from
the European Commission’s Framework Programme 7, through the Marie Curie
International Research Staff Exchange Scheme LACEGAL (PIRSES–GA–2010–269264).
LAB acknowledges support from CONICYT FONDECYT/POSTDOCTORADO/3180359. PBT
acknowledges partial support from Proyecto Interno UNAB: Galaxy Formation and
Chemical Evolution.




\bibliographystyle{mnras}
\bibliography{EagleMorph} 



\appendix

\section{Dependence on orientation}
\label{sec:dependence_on_rotation}

\begin{figure*}
    \includegraphics[width=\columnwidth]{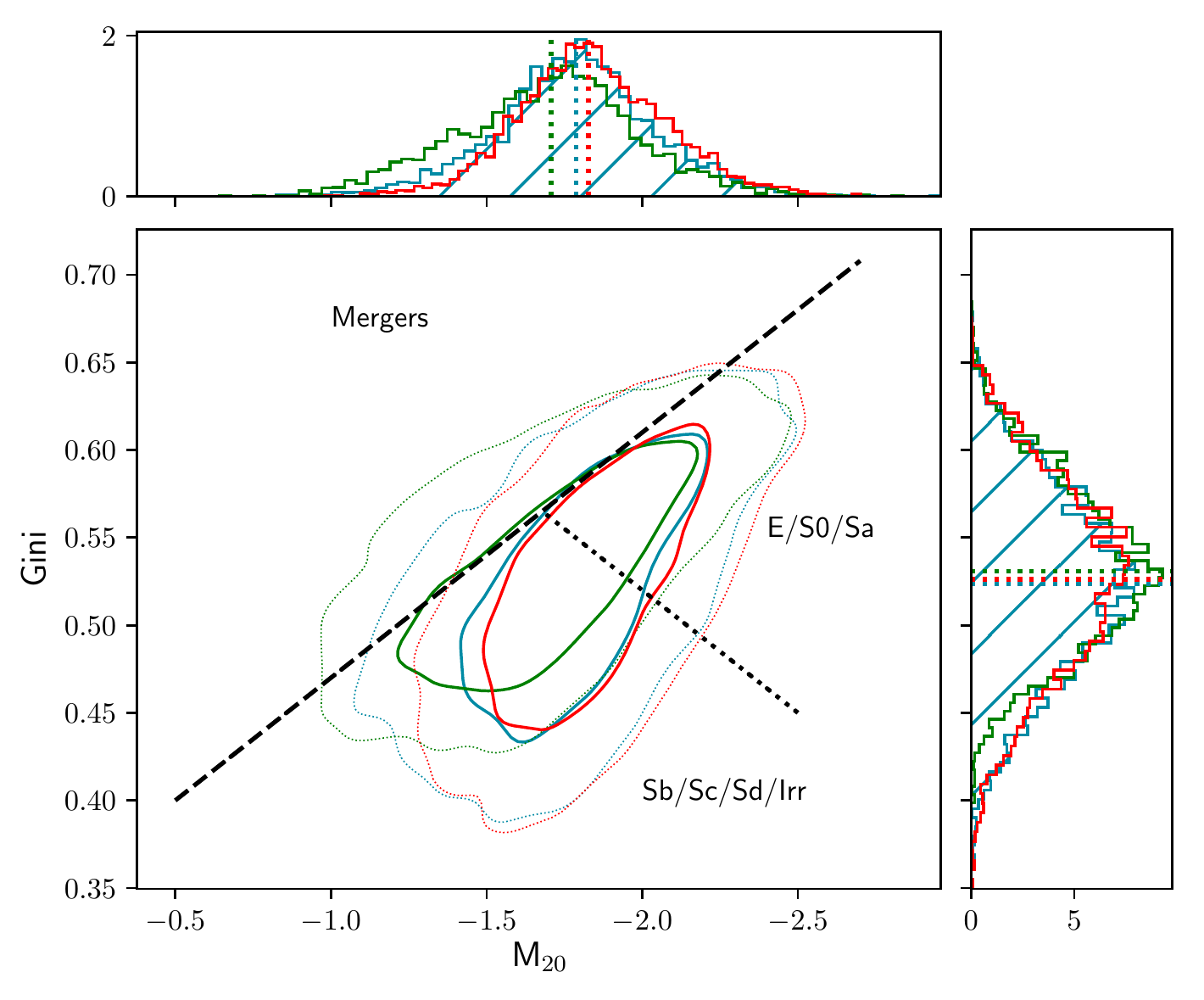}
    \includegraphics[width=\columnwidth]{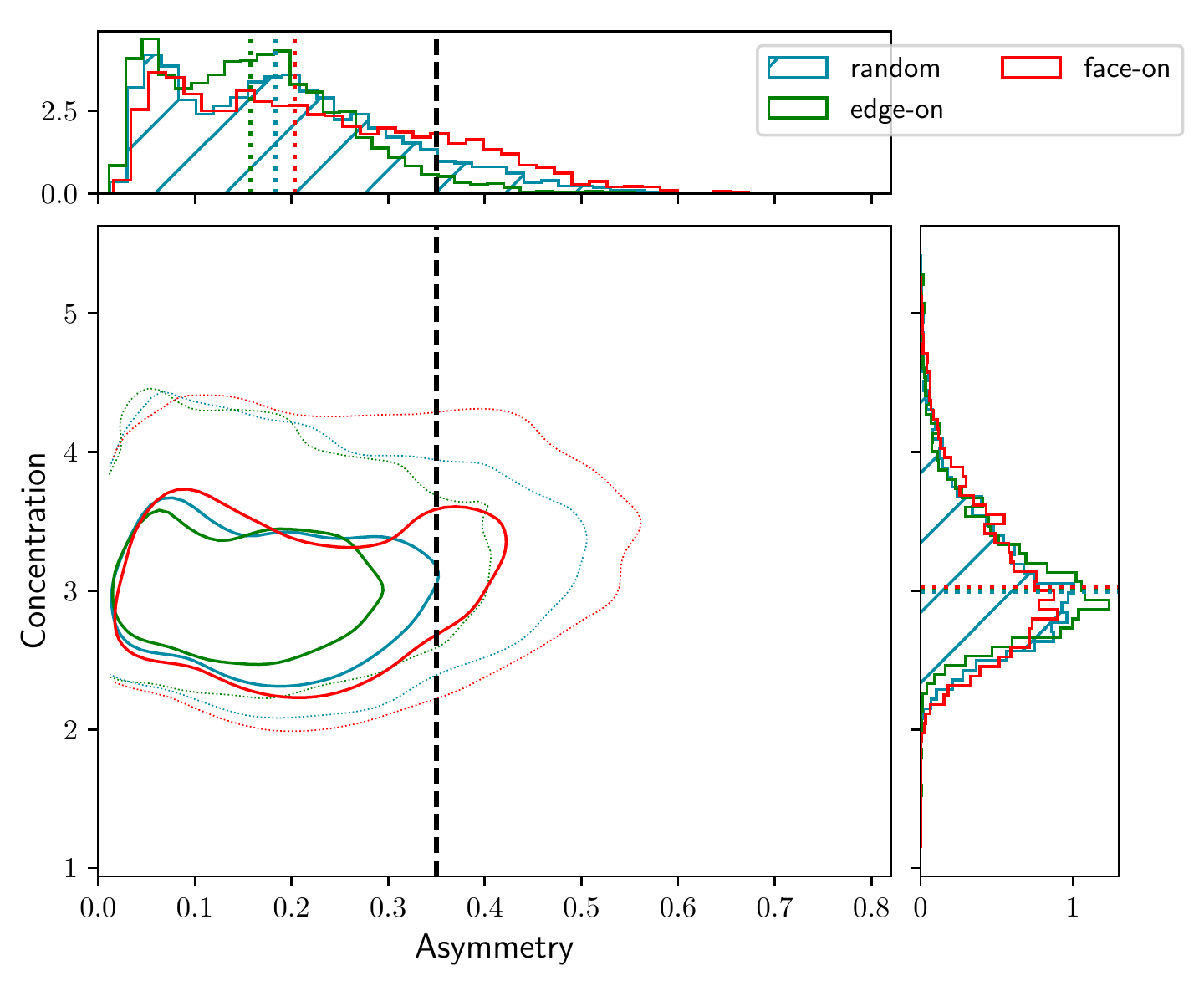}
    
    \caption{$G$ versus M$_{20}$ (panel on the left) and $C$ versus
        $A$ (panel on the right) for Ref-100 galaxies. We compare the
        results of three viewing angles: face-on, edge-on and a random
        orientation with respect to the galaxy (but fixed to the $xy$ plane of
        the simulation box). The random orientation was used thought the main
        body of the paper.}
    
    \label{fig:scatter_orientations}
\end{figure*}

In the main body of the paper we present the results corresponding to a random
orientation with respect to the galaxy (but fixed to the $xy$ plane of the
simulation box). Here we compare those results with two other extreme viewing
angles: edge- and face-on views. Figure \ref{fig:scatter_orientations} shows the
variation in the distributions of $G$, M$_{20}$, $C$ and $A$
caused by considering the different viewing angles. Compared to their randomly
oriented counterparts, we find that edge-on views result in higher median
M$_{20}$ and $G$ values (4\% and 1.3\%, respectively). However, the largest
changes are found for the case of $A$ where edge-on (face-on) views result
in a median value 14\% (10\%) lower (higher). Unsurprisingly, the effects of
orientation are more noticeable in lower-mass, disky and star-forming galaxies.

\section{Numerical convergence}
\label{sec:numerical_convergence}

\begin{figure*}
    \includegraphics[width=\columnwidth]{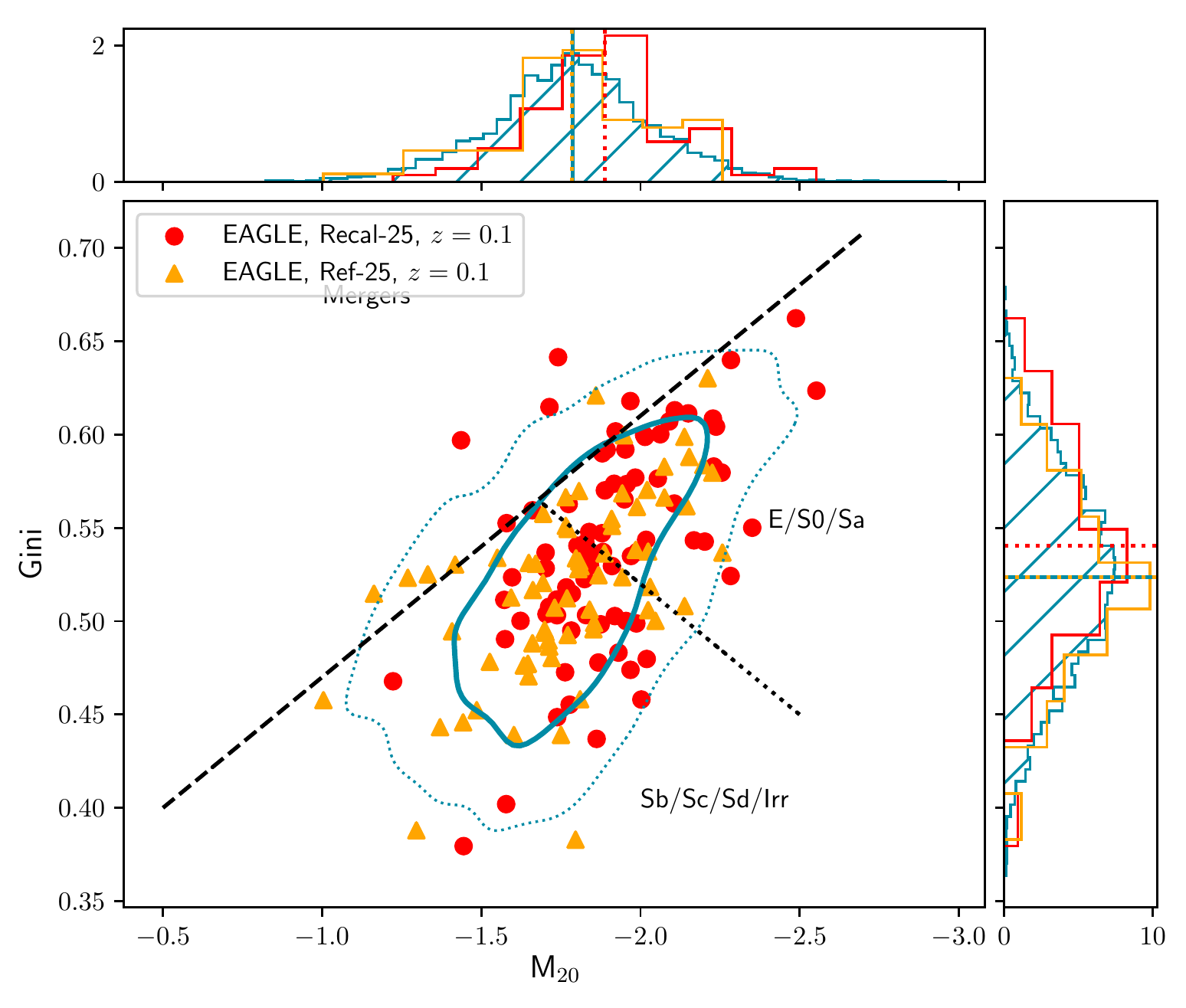}
    \includegraphics[width=\columnwidth]{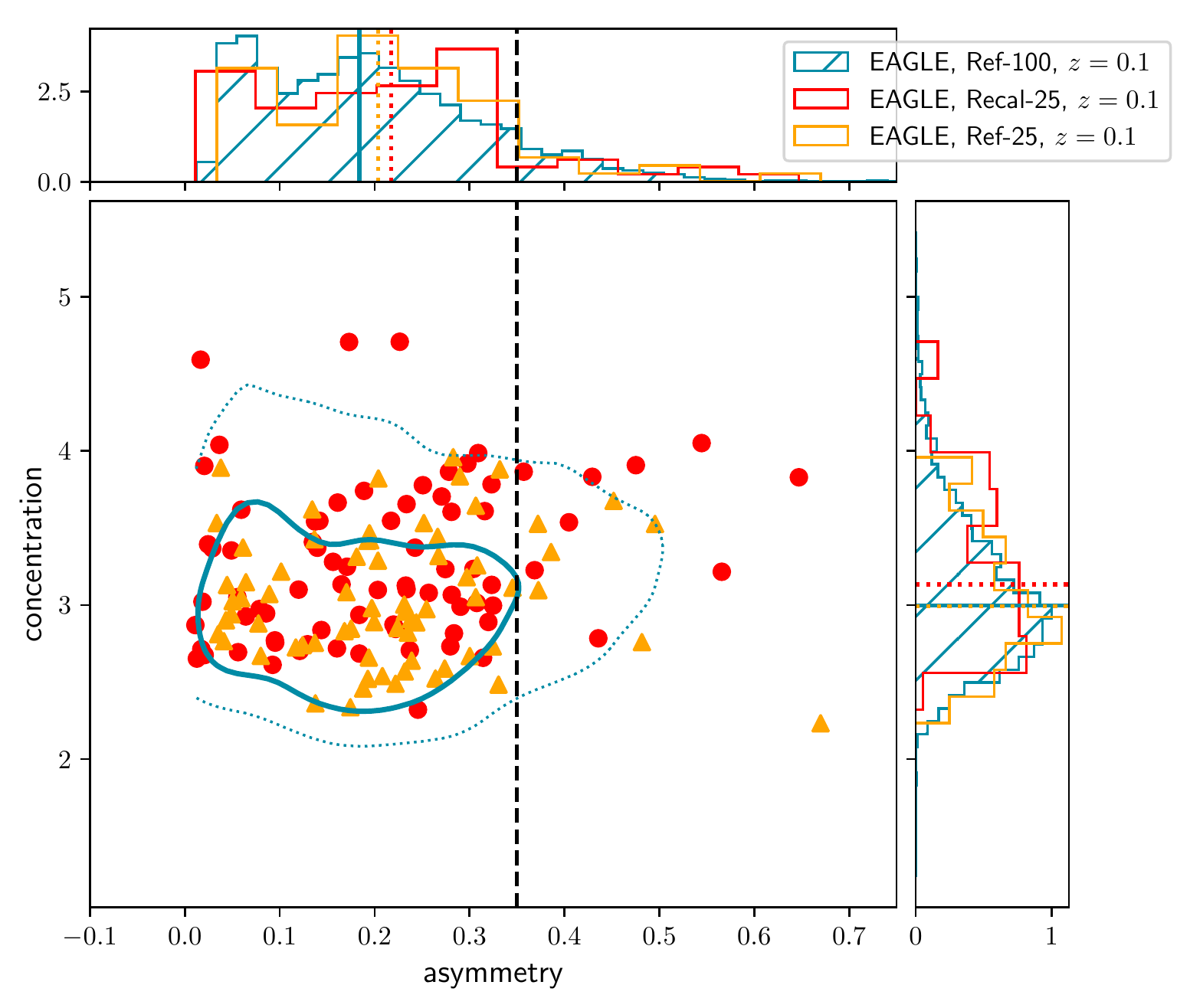}
    
    \caption{$G$ versus M$_{20}$ (panel on the left) and $C$ versus
    $A$ (panel on the right) for galaxies in the Recal-25 (red points and
    lines), Ref-25 (orange triangles and lines) and Ref-100 (cyan contours)
    simulations. Non-parametric statistics were computed from a face-on viewing
    angle.}
    
    \label{fig:scatter_sims}
\end{figure*}

To test numerical convergence, we compare the non-parametric morphologies of
galaxies in the Ref-100 simulation with those in a simulation with resolution a
factor 8 finer in mass and a factor 2 finer in length scale, Recal-25. This
constitutes a test of {\it weak convergence}, as discuss in
\citet{schaye_eagle_2015}, given that the subgrid model of the higher resolution
simulation has been recalibrated. In Figure \ref{fig:scatter_sims} we show the
distributions of $G$, M$_{20}$, $C$ and $A$ for galaxies in
Ref-100 (cyan lines), Recal-25 (red points and lines) and Ref-25 (orange
triangles and lines). Non-parametric statistics were computed from a face-on
viewing angle, to eliminate the effects of orientation. We find that Recal-25
galaxies occupy a similar distribution in morphological space as its lower
resolution counterparts. However, there are slight tensions.

In table \ref{tab:ks_numerical} we show the results of two-sample
Kolmogorov–Smirnoff (KS) tests comparing the distributions of morphological
statistics of galaxy samples extracted from Recal-25 against the sample
extracted from the Ref-100 simulation. We find that the null hypothesis that
Recal-25 and Ref-100 share the same distribution of morphological parameters
should be marginally rejected for the cases of M$_{20}$ and $C$.
Galaxies in Recal-25 exhibit a median M$_{20} $($C$) 3.3\% (3.5\%)
higher (lower) than galaxies in Ref-100. The other statistics considered: $G$ and
$A$ have distributions that are statistically equivalent according to the KS
tests.

Ref-25 has the same volume size as Recal-25, but shares resolution with Ref-100
instead, its inclusion serves to control against box-size effects. The same
KS-test applied to the Ref-25 simulation indicate that the simulation share
similar distributions with Ref-100 for all statistics considered, which allow us
to reject effects due to the smaller volume. 

The differences between Recal-25 and Ref-100 are not large enough to affect the
morphological trends presented in the main body of the paper and therefore do
not affect our conclusions significantly. It is also interesting to point out
that the mock morphological statistic that most diverges from the observations,
$A$, is not particularly affected by numerical resolution. Therefore we
should look for the reasons of this discrepancy in the image generation
procedures.

\begin{table}
	\centering
	\caption{Results of two-sample Kolmogorov–Smirnoff tests comparing galaxy
        samples extracted from Recal-25 and Ref-25 against a sample extracted
        from the Ref-100 simulation}
	\label{tab:ks_numerical}
    \begin{tabular}{l l l l l}
        \hline
              & \multicolumn{2}{c}{Recal-25} & \multicolumn{2}{c}{Ref-25} \\
              & D & p-value & D & p-value \\
        \hline                                                                                                                     
        $G$        & 0.13 & 0.12   & 0.13 & 0.18 \\
        M$_{20}$       & 0.16 & 0.04   & 0.13 & 0.16 \\
        $C$   & 0.18 & 0.01   & 0.09 & 0.65 \\
        $A$       & 0.13 & 0.16   & 0.12 & 0.29 \\
        \hline
        
    \end{tabular}
\end{table}


\bsp	
\label{lastpage}
\end{document}